\documentclass{aamas2016}

\usepackage{times}
\usepackage{rotating}
\usepackage{verbatim}
\usepackage{amssymb}
\usepackage{amsmath}
\usepackage{array}
\usepackage{float}
\usepackage{graphics,graphicx,epsfig,epstopdf}
\usepackage{psfrag}
\usepackage{subfigure}
\usepackage{tikz}
\usepackage{url}
\usepackage{color}
\usepackage{theorem}
\usepackage{algorithm}
\usepackage{algorithmic}
\usepackage{multirow}
\usepackage{mathtools}
\usepackage{paralist}

\usepackage{hyperref}
\hypersetup{
    bookmarks=true,         
    unicode=false,          
    pdftoolbar=true,        
    pdfmenubar=true,        
    pdffitwindow=false,     
    pdftitle={},    
    pdfauthor={},     
    pdfsubject={},   
    pdfcreator={},   
    pdfproducer={},  
    pdfkeywords={}{}{}, 
    pdfnewwindow=true,      
    colorlinks=false,       
    linkcolor=red,          
    citecolor=green,        
    filecolor=magenta,      
    urlcolor=cyan           
}

\pdftrailerid{}
\newtheorem{proposition}{Proposition}

\newcommand{\iin}{\! \in \!}

\newcommand*{\QEDB}{\hfill\ensuremath{\square}}%

\begin{document}

\title{Methods for finding leader--follower equilibria with multiple followers}

\numberofauthors{3}

\author{
\alignauthor
Nicola Basilico\\
       \affaddr{University of Milan}\\
       \affaddr{Milan, Italy}\\
       \email{nicola.basilico@unimi.it}
\alignauthor
Stefano Coniglio\\
       \affaddr{University of Southampton}\\
       \affaddr{Southampton, UK}\\
       \email{s.coniglio@soton.ac.uk}
\alignauthor
Nicola Gatti\\
       \affaddr{Politecnico di Milano}\\
       \affaddr{Milan, Italy}\\
       \email{ngatti@elet.polimi.it}
}

\maketitle

\begin{abstract}
The concept of leader--follower (or Stackelberg) equilibrium plays a central role in a number of real--world applications of game theory.
While the case with a single follower has been thoroughly investigated, results with multiple followers are only sporadic and the problem of designing and evaluating computationally tractable equilibrium-finding algorithms is still largely open.
In this work, we focus on the fundamental case where multiple followers play a Nash equilibrium once the leader has committed to a strategy---as we illustrate, the corresponding equilibrium finding problem can be easily shown to be $\mathcal{FNP}$--hard and not in Poly--$\mathcal{APX}$ unless $\mathcal{P} = \mathcal{NP}$ and therefore it is one among the hardest problems to solve and approximate. 
We propose nonconvex mathematical programming formulations and global optimization methods to find both exact and approximate equilibria, as well as a heuristic black box algorithm. 
All the methods and formulations that we introduce are thoroughly evaluated computationally.
\end{abstract}




\section{Introduction}
The design of computationally viable techniques for the construction of game--theoretical solutions to real--life problems has recently become a central topic in Artificial Intelligence.
Besides several classical solution concepts, such as those of Nash Equilibrium (NE) and Correlated Equilibrium (CE)--- see~\cite{shoham-book} for a thorough exposition---, the concept of \emph{Leader--Follower Equilibrium} (LFE) has received the arguably largest share of attention, thanks to its many applications, especially in the security domain~\cite{an2011guards,DBLP:conf/atal/KiekintveldJTPOT09}.

The computational study of LFEs is well assessed for the case of a single follower who, in an equilibrium, is known to play w.l.o.g. a pure strategy~\cite{leaderfollower}. 
In this case, the problem of computing an LFE is easy with complete information, whereas it is $\mathcal{FNP}$--hard for Bayesian games~\cite{Conitzer:2006:COS:1134707.1134717}. 
Solution algorithms are proposed in~\cite{Conitzer:2006:COS:1134707.1134717}.
With multiple followers, different cases are possible depending on the nature of the game they play. 
Applications can be found in, among others, {\em social planning} (e.g., urban traffic plans and economic markets), {\em monetary economics} (e.g., quantitative easing by the European Central Bank), and {\em security} (e.g., NATO protection of civilians in conflicts between more armies).
Only sporadic results are available in the literature. 
Among them, it is known that, if the followers play a CE, an LFE can be found in polynomial time~\cite{DBLP:conf/aaai/ConitzerK11} whereas, if they play sequentially one at a time (as in a classical Stackelberg game), the problem is $\mathcal{FNP}$--hard~\cite{Conitzer:2006:COS:1134707.1134717}.

In this paper, we focus on the fundamental case of leader--follower games with multiple followers where the latter play simultaneously in a noncooperative way, thus playing an NE for any commitment of the leader. 
We refer to the corresponding LFE as {\em Leader--Follower Equilibrium Nash} (LFE--N).

The original contributions of our work are as follow.
We provide, to the best of our knowledge, the first exact and approximate methods to compute an LFE--N.
We illustrate how computing an LFE--N when the followers play an NE which either maximizes ({\em optimistic} case) or minimizes ({\em pessimistic} case) the leader's utility can be easily shown to be $\mathcal{FNP}$--hard and not in Poly--$\mathcal{APX}$ unless $\mathcal{P} = \mathcal{NP}$ (i.e., it is unlikely that there is a polynomial--time algorithm approximating the optimal value within an approximation ratio that depends polynomially on the size of the game), even for polymatrix games. 
Thus, this problem results to be among the hardest problems to solve and approximate.
After highlighting the clear bilevel nature of the problem, we propose different nonlinear (and nonconvex) mathematical programs for the optimistic case, resorting to {\em global optimization} tools that to the best of our knowledge, have not yet been thoroughly applied in algorithmic game theory.
For the {\em pessimistic} case, where well--established techniques to cast the corresponding bilevel program into a compact single level one do not apply, we propose a method based on the combination of global optimization and \emph{black box optimization} techniques. 
We also tailor our methods for polymatrix games, which play an important role in leader--follower scenarios, e.g., in security domains where the defender (acting as the leader) may need to optimize against multiple uncoordinated attackers (acting as the followers). 
We provide a thorough experimental evaluation of our methods on a standard (normal--form and polymatrix) testbed generated with GAMUT~\cite{gamut}, employing different {\em global optimization} solvers---BARON~\cite{BARON} and SCIP~\shortcite{SCIP}, based on {\em spatial branch--and--bound}, and CPLEX~\shortcite{CPLEX}, based on {\em branch--and--bound}---and {\em local optimization solvers}---RBFOpt~\cite{blackboxoptimization}, based on {\em black box optimization}, and SNOPT~\cite{Gill97snopt:an}, based on {\em sequential quadratic programming}.  
We show that our global optimization methods solve exactly game instances with a size (e.g., with three players, 9 and 15 actions per player in normal--form and polymatrix games, respectively) similar to that of the largest instances solved by state--of--the--art NE finding algorithms (less than 10 actions with three players~\cite{DBLP:journals/geb/PorterNS08}), while they provide very good approximations (with a multiplicative gap smaller than 35\% for normal--form games and 30\% for polymatrix games) up to instances with three players and more than 40 and 45 actions per player in normal--form and polymatrix games, respectively, corresponding to more than 64,000 different outcomes.

\section{Preliminaries and complexity}
Let $N=\{1,\dots,n\}$ be a set of agents and, for each $i \in N$, let $A_i$ be the corresponding set of actions, with $m_i = |A_i|$. 
For each agent $i \in N$, we denote by $x_i \in [0,1]^{m_i}$, with $e^T x_i = 1$ (where $e$ is the all--one vector), his {\em strategy vector} (or strategy, for short), where each component $x_{i}^a$ represents the probability by which agent~$i$ plays action~$a \in A_{i}$. 
We denote a {\em strategy profile}, i.e., the collection of the strategies of the different agents, by $x=(x_{1}, \ldots, x_{n})$. Let $u_i(x_1, \dots, x_n)$ be the expected utility of agent~$i \in N$. 
A strategy profile $x=(x_{1}, \ldots, x_{n})$ is an NE if and only if, for each agent $i\in N$, $u_i(x_{1},\ldots,x_{n}) \geq u_i(x'_{1},\dots,x'_{n})$ for any strategy profile $x'$ where $x'_j=x_j$ for all $j \in N \setminus \{i\}$ and $x'_i \neq x_i$ (no unilateral deviation).
We consider two game classes: {\em Normal--Form} (NF) and {\em PolyMatrix} (PM).

For NF games~\cite{shoham-book}, let $U_i \in \mathbb{R}^{m_1 \times \ldots \times m_n}$ denote, for each agent $i \in N$, his  (multidimensional)  utility (or payoff) matrix, where each component $U_i^{a_1,\ldots,a_n}$ denotes the utility of agent~$i$ when all the agents play actions $a_1,\ldots,a_n$. 
Given a strategy profile $(x_1, \ldots, x_n)$, the expected utility of agent~$i \in N$ is defined by the multilinear function $u_i(x_{1},\ldots,x_{n}) =  x_i^T \left(U_{i}\cdot \prod_{j \in N \setminus\{i\}}x_{j}\right)$ (an $n$th--degree polynomial).

For PM games~\cite{polymatrixref}, we have a matrix $U_{ij} \in \mathbb{R}^{m_i \times m_j}$ per pair of agents $i,j \in A$. 
Given a strategy profile $(x_1, \ldots, x_n)$, the expected utility of agent~$i$ is defined as the bilinear function $u_i(x_{1},\ldots,x_{n})= \sum_{j \in N \setminus\{i\}} x_{i}^T U_{ij} x_{j}$ (a 2nd--degree polynomial).

By virtue of the correctness of the mathematical programming formulations that we will propose, an LFE--N is guaranteed to exist when the followers maximize the leader's expected utility (optimistic case). 
Differently, it may not exist in the pessimistic case, as it is known for the case with a single follower~\cite{leaderfollower}.

It can be shown that computing an optimistic or a pessimistic LFE--N is $\mathcal{FNP}$--hard and, unless $\mathcal{P}=\mathcal{NP}$, does not admit any polytime approximation algorithm with a ratio that is polynomial in the size of the game.
Due to our paper being mainly experimental, we discuss this result, which is necessary to characterize the hardness of the problem, in Appendix~A.
Furthermore, as we discuss in Appendix~A, it can be shown that the problem of deciding whether an action of the leader can be safely discarded (because always played with 0 probability in an LFE--N) is $\mathcal{NP}$--hard.
In this work, we mostly focus on the most general case where both the leader and the followers play mixed strategies. 
We also tackle the case where the leader plays only pure strategies, as it allows for more efficient algorithms\footnote{\scriptsize The case where leader and followers play only pure strategies is trivial in both optimistic and pessimistic versions, as it can be solved in $O(m^n)$ by enumeration, as much as the case where the leader plays mixed strategies and the followers are allowed to play only pure strategies in the optimistic version, which is solved by solving $O(m^{n-1})$ linear programs. The pessimistic version of the latter can be tackled with the same techniques that we propose for the fully mixed case.}.

\section{Algorithms and methods}
In the following, we assume that the $n$th agent, whom we relabel as agent~$\ell$, takes the leader's role. We denote the other agents (the followers) by the set $F = N \setminus \{\ell\}$.
For the sake of presentation, we present our formulations for $n=3$ (one leader, two followers) although they can be easily adapted to any $n$ (see our computational experiments for up to $n = 6$). 
Let thus $F = \{1,2\}$. For all $f \in F$, we will adopt the notation $f' := F \setminus \{f\}$. We will also denote~$x_\ell$ by~$\delta$ and $x_{1}, x_{2}$ by $\rho_1,\rho_2$.

The computation of an LFE--N amounts to solving a {\em bilevel program}. In the first level, we look for a strategy~$\delta$ while, in the second level and for the given $\delta$, we look for two strategies $\rho_1,\rho_2$ forming an NE which either maximizes (optimistic case) or minimizes (pessimistic case) the leader's utility. In the general case (under mild technical assumptions), if we assume the convexity of the second level problem, bilevel programs can be cast as (compact) single level mathematical programs by substituting for the second level problem its KKT conditions~\cite{dempe2003bilevel}.

For the {\em Optimistic} (O) version of LFE--N, the optimality conditions are not needed, as we can turn the second level problem into one of pure feasibility over which the leader has full control: the leader looks for a strategy vector $\delta$ and, given~$\delta$, also for an NE in the follower's game such that his utility is maximized. As we will show, this allows us to solve the problem exactly via (nonlinear) mathematical programming.
For the {\em Pessimistic} (P) version of LFE--N, we clearly cannot get rid of the second level objective function as the leader cannot control which NE the followers choose. Moreover, KKT conditions do not yield a compact reformulation, as even the sole feasible region of the second level problem (which corresponds to the set of NEs of a game, parameterized by $\delta$) is highly nonconvex~\cite{shoham-book}. For this case, we resort to (heuristic) black box optimization techniques, assuring only a lower bound on the leader's utility in an optimal LFE--N.

\subsection{Leader in Mixed strategies and Followers in Mixed stragies (LMFM)}
We first focus on the optimistic case for NF and PM games. We propose three different exact mathematical programming formulations for NF games and illustrate how they simplify for PM games. We then address the pessimistic case, proposing a black box approach with an exact mathematical programming oracle.

\subsubsection{Exact formulations for Optimistic Normal Form LMFM games (O--NF--LMFM)}

\vspace{-0.2cm}

\paragraph{O--NF--LMFM--I} Let, for each $f \in F$, $v_f$ be his {\em best response} value. We start with a formulation obtained by casting the second level problem as a Linear Complementarity Problem (LCP) which, as the followers play an NF game parameterized by the leader's strategy $\delta$, becomes, rather than linear, bilinear for $n=3$:

\begin{scriptsize}
\begin{align}\label{first-begin}
\max \quad \sum_{i \in A_\ell}\sum_{j \in A_1}\sum_{k \in A_2} \delta^i \rho_{1}^j\rho_{2}^k U_\ell^{ijk}
  & \quad\text{s.t.}\\
v_f \geq \sum_{i \in A_\ell} \sum_{k \in A_{f'}} \delta^i \rho_{f'}^k {U}_f^{ijk}
  & \quad \forall f \in F, j \in A_f\\
\sum_{j \in A_f} \rho_f^j \big(v_f - \sum_{i \in A_\ell} \sum_{k \in A_{f'}} \delta^i \rho_{f'}^k {U}_f^{ijk}\big) = 0 
  & \quad\forall f \in F\label{nonlincompl}\\
\sum_{i \in A_\ell}\delta^i = 1, \delta \geq 0  \label{delta}\\
\sum_{j \in A_f} \rho_f^j = 1, \rho_f \geq 0 & \quad \forall f \in F. \label{rho}
\end{align}
\end{scriptsize}

\noindent The problem contains $|F|=2$ cubic constraints, $m_1+m_2$ quadratic constraints, and a cubic objective function.

\paragraph{O--NF--LMFM--II} For each $f \in F$, let $u_f^j$ be the utility agent $f$ expects when playing action $j \in A_f$ and let~$r_f^j$ be the corresponding {\em regret value}. Let $M_f=\max\limits_{i \in A_\ell, j \in A_1, k \in A_2}\{U_f^{ijk}\} - \min\limits_{i \in A_\ell, j \in A_1, k \in A_2}\{U_f^{ijk}\}$. Adopting a reformulation similar to that in~\cite{sandholmgilpinconitzer2005}, we remove the LCP constraints by introducing, for each follower $f \in F$, a binary vector of variables~$s_f$ corresponding to the {\em support} of the strategy vector~$\rho_f$, so that, for any $j \in A_f$, $s_f^j = 1 \Rightarrow \rho_f^j = 0$. We have:

\vspace{-.1cm}
\begin{scriptsize}
\begin{align}
\label{one} \hspace{-0.5cm}
\max \sum_{i \in A_\ell}\sum_{j \in A_1}\sum_{k \in A_2} \delta^i\rho_1^j\rho_2^k {U}_\ell^{ijk}
  & \quad \text{s.t.} \\
\label{two}
     {u}_f^j = \sum_{i \in A_\ell} \sum_{k \in A_{f'}}  \delta^i\rho_{f'}^k  {U}_f^{ijk}
    &  \quad \forall f\in F, j \in A_f\\
\label{cons1}
     {v}_f \geq {u}_f^j
    & \quad \forall f\in F, j \in A_f\\
\label{cons2}
     r_f^j = v_f- {u}_f^{j}
    & \quad \forall f\in F, j \in A_f\\
\label{cons3}
     \rho_f^j \leq 1 - s_f^{j}
  & \quad \forall f\in F, j \in A_f\\
\label{cons4}
     r_f^j \leq M_f s_f^{j}
  & \quad \forall f \in F, j \in A_f\\
\label{consSupp}
    s_f^j \in \{0,1\}
  & \quad \forall f \in F, j \in A_f\\
\textnormal{Constraints (\ref{delta})--(\ref{rho}).}
\end{align}
\end{scriptsize}

\noindent With this formulation we achieve, at the cost of introducing binary variables, fewer nonlinearities: only $m_1+m_2$ quadratic constraints and a cubic objective function.

\paragraph{O--NF--LMFM--III} This third formulation is obtained from O--NF--LMFM--II by first carrying out the reformulation steps that are performed in a standard spatial branch--and--bound algorithm (employed to solve a nonlinear program to global optimality), and then tightening the resulting formulation via valid constraints. We restate each original multilinear term by introducing (iteratively) an auxiliary variable and a bilinear constraint, as in Constraints~\eqref{zik}--\eqref{zijk}. We obtain:

\begin{scriptsize}
\begin{align}
\max \quad \sum_{i \in A_\ell}\sum_{j \in A_1}\sum_{k \in A_2} z^{ijk}{U}_{\ell}^{ijk} & \quad \text{s.t.} \label{objectino}\\
u_f^j = \sum_{i \in A_\ell} \sum_{k \in A_{f'}}  y_{f'}^{ik}  {U}_f^{ijk} & \quad \forall f\in F, j \in A_f \label{cons22}\\
y^{ij}_f = \delta^i \rho_f^j & \quad \forall i \in A_\ell, f \in F, j \in A_f \label{zik}\\
z^{ijk} = y_1^{ij} \rho^k_2 & \; \forall i \iin A_\ell,  j \iin A_1, k \iin A_{2} \hspace{-1cm}\label{zijk}\\
\sum_{i \in A_\ell}\sum_{j \in A_{f}} y_{f}^{ij} = 1 & \quad \forall f \in F \label{cons5} \\ 
\sum_{i \in A_\ell}\sum_{j \in A_1}\sum_{k \in A_2} z^{ijk} = 1\label{cons6} \\
y^{ij}_f \geq 0 & \quad \forall f \in F, i \in A_{\ell}, j \in A_f \hspace{-1cm}\label{cons7}\\
z^{ijk} \geq 0 & \quad \forall i \iin A_\ell,  j \iin A_1, k \iin A_{2} \label{cons8}\\
\textnormal{Constraints \eqref{delta}--\eqref{rho}, \eqref{cons1}--\eqref{consSupp}.} \label{finalino}
\end{align}
\end{scriptsize}

\noindent The advantage of carrying out this reformulation step {\em a priori} is that, when explicitly introducing the variables $y^{ij}_{f}$ and $z^{ijk}$ representing, resp., the products $\delta^i\rho^j_f$ and $\delta^i\rho^j_1\rho^k_2$, we can tighten the new formulation. Indeed, for any $x,y \in \mathbb{R}^n$, the linear equations $e^T x = e^T y = 1$ imply the validity of $e^T (x y^T) e = (e^T x) (y^T e) = 1$, which translates into Equations~\eqref{cons5} and~\eqref{cons6}.
Overall, we obtain $m_\ell (m_1 + m_2) + m_\ell m_1 m_2$ quadratic constraints and a linear objective function, yielding a tighter formulation.

\subsubsection{Exact formulations for Optimistic Polymatrix LMFM games (O--PM--LMFM)}

For PM games, for any $f \in F$, the expected utility $u_f^j$ for action $j \in A_f$ (which is trilinear for NF games with $n=3$, and of order $n$ in general) is defined as the linear (for any $n$) function $u_f^j = \sum_{i \in A_\ell} \delta^i {U}_{f\ell}^{ij} + \sum_{k \in A_{f'}} \rho^{k}_{f'} {U}^{jk}_{ff'}$. The leader's utility is the bilinear (for any $n$) function $\sum_{i \in A_\ell} \sum_{f \in F}\sum_{j \in A_f} \delta^i \rho_f^j U_{\ell f}^{ij}$. As a consequence, the PM counterparts to Formulations I, II, and III contain, in general, fewer nonlinearities. Indeed, O--PM--LMFM--I only contains $|F|=2$ quadratic constraints and a quadratic objective (as Constraints~(2)--(3) and Objective~(1) become, resp., linear, quadratic, and quadratic).
O--PM--LMFM--II contains only linear constraints, binary variables, and a quadratic objective (as Constraints~(7) and Objective~(6) become, resp. linear and quadratic).
O--PM--LMFM--III contains only $m_\ell (m_1+m_2)$ quadratic constraints, binary variables, and a linear objective function. The latter is derived, similarly to O--NF--LMFM--III, by reformulation of each multilinear term in O--PM--LMFM--II; since, in the latter, the only nonlinearity is in the objective, O--PM--LMFM--III is obtained by just reformulating the products $\delta^i \rho_f^j$ it contains, for all $f \in F$ and $j \in A_f$ and adding the counterpart to Constraints~(19). The three formulations read as follows.

\subsubsection*{O--PM--LMFM--I}
\begin{scriptsize}
\begin{align}   
\hspace{-0.6cm}\max \quad \sum_{i \in A_\ell} \sum_{f \in F}\sum_{j \in A_f} \delta^i \rho_f^j U_{\ell f}^{ij} 																& 	~~\text{s.t.}														 \\
v_f \geq \sum_{i \in A_\ell} \delta^i U_{f\ell}^{ij} + \sum_{k \in A_{f'}} \rho^{k}_{f'} U_{ff'}^{jk} & \quad \forall f \in F, j \in A_f\\
\hspace{-0.7cm}\sum_{j \in A_{f}} \rho_f^j \big(v_f - \sum_{i \in A_\ell} \delta^i U_{f\ell}^{ij} - \sum_{k \in A_{f'}} \rho^{k}_{f'} U_{ff'}^{jk}\big) = 0 & \quad \forall f \in F\\
\textnormal{Constraints (\ref{delta})--(\ref{rho})}
\end{align}
\end{scriptsize}

\subsubsection*{O--PM--LMFM--II} 

\begin{scriptsize}
\begin{align}
\label{myinit}\hspace{-0.5cm}\max  \quad \sum_{i \in A_\ell} \sum_{f \in F}\sum_{j \in A_f} \delta^i \rho_f^j U_{\ell f}^{ij}									  			& 	\quad \text{s.t.} 																\\
{u}_f^j = \sum_{i \in A_\ell} \delta^i U_{f\ell}^{ij} + \sum_{k \in A_{f'}} \rho^{k}_{f'} U_{ff'}^{jk} & \quad \forall f \in F, j \in A_f\label{cons100}	\\
\textnormal{Constraints \eqref{delta}--\eqref{rho}, \eqref{cons1}--\eqref{consSupp}}
\end{align}
\end{scriptsize}


\subsubsection*{O--PM--LMFM--III} 

\begin{scriptsize}
\begin{align}
\hspace{-0.5cm}\max \quad \sum_{i \in A_\ell} \sum_{f \in F}\sum_{j \in A_f} y_{f}^{ij}  U_{\ell f}^{ij} & \quad\text{s.t.} \\
{u}_f^j = \sum_{i \in A_\ell} \delta^i U_{f\ell}^{ij} + \sum_{k \in A_{f'}} \rho^{k}_{f'} U_{ff'}^{jk} & \quad \forall f \in F, j \in A_f\\
y_{f}^{ij} = \delta^i \rho^j_f & \quad \forall i \in A_\ell, f \in F, j \in A_f\\
\sum_{i \in A_\ell} \sum_{j \in A_f} y_{f}^{ij} = 1 & \quad \forall i \in A_\ell, f \in F\\
y^{ij}_f \geq 0 & \quad \forall i \in A_\ell,  f \in F, j \in A_f\\
\textnormal{Constraints \eqref{delta}--\eqref{rho}, \eqref{cons1}--\eqref{consSupp}}
\end{align}
\end{scriptsize}



\subsubsection{Black box method for optimistic and pessimistic normal form or polymatrix LMFM games}

For both the optimistic and pessimistic cases, we propose a black box approach based on a Radial Basis Function (RBF) estimation, relying on the solver RBFOpt~\cite{blackboxoptimization}. The idea is of exploring the leader's strategy space (variables $\delta$) with a direct search that, iteratively, builds an RBF approximation of the objective function, relying on the solution of an {\em oracle formulation} for the objective function evaluation.
Given any incumbent value $\hat \delta$, the oracle solves the (NF or PM) optimistic or pessimistic second level problem exactly via one of our formulations, after imposing $\delta = \hat \delta$.
For optimistic NF games, we propose an oracle formulation similar to O--NF--LMFM--III, obtained from O--NF--LMFM--II by adopting a different reformulation with auxiliary variables $y^{jk} = \rho_1^j \rho_2^k$, which is tighter than that obtained from O--NF--LMFM--III when $\delta$ is given. It only contains a quadratic objective, linear constraints, and binary variables. For PM games, the oracle formulation is a Mixed--Integer Linear Program (MILP). For the pessimistic cases, the leader's utility is minimized. Notice that, in this last case, we might search for an equilibrium that does not exists. Nevertheless, our method would return an approximate solution in any case since the solution space we explore (leader's strategy space) is finite. For a given $\delta$, the two oracle formulation read as follows (we report only the optimistic versions, the pessimistic counterparts can be obtained by simply replacing $\max$ with $\min$ in the objective function).

The formulation reads:
\subsubsection*{O/P--NF/PM--LMFM--BlackBox, oracle formulation}

\begin{scriptsize}
\begin{align}
\max \quad \sum_{i \in A_\ell}\sum_{j \in A_1}\sum_{k \in A_2} \delta^i y^{jk}{U}_{\ell}^{ijk} & \quad \text{s.t.}\\
u_f^j = \sum_{i \in A_\ell} \sum_{k \in A_{f'}}  \delta^i \rho_{f'}^{k}  {U}_f^{ijk} & \quad \forall f\in F, j \in A_f\\
y^{jk} = \rho_1^j \rho^k_2 & \quad \forall j \in A_1, k \in A_{2} \hspace{-1cm}\\
\sum_{j \in A_1}\sum_{k \in A_2} y_{f}^{jk} = 1\\ 
y^{jk} \geq 0 & \quad \forall j \in A_1, k \in A_2 \hspace{-1cm}\\
\textnormal{Constraints~\eqref{rho}, \eqref{cons1}--\eqref{consSupp}.}
\end{align}
\end{scriptsize}



\subsection{Leader in pure strategies and followers in mixed strategies (LPFM)}

We focus on the case in which the leader is restricted to pure strategie. This case is of interest when the followers can see the action actually played by the leader and, therefore, the leader cannot commit to a mixed strategy. Here, we propose
an {\em ad hoc} implicit enumeration algorithm, more efficient than solving the previously proposed formulations. Since a solution can be found by solving any of our formulations after imposing $\delta \in \{0,1\}^{m_\ell}$, and the LMFM counterpart to O--NF--LPFM--III contains, as we will see, fewer nonlinearities than the original one, we report it for comparisons.

\subsubsection{O--NF/PM--LPFM--Implicit--Enumeration}
Due to $\delta \in \{0,1\}^{m_\ell}$, an LFE--N can be found by solving $m_\ell$ times one of our formulations, iteratively fixing $\delta = e_i$ (where $e_i$ is the all zero vector with a single 1 in position $i$), and selecting the best outcome. The idea of the algorithm is of pruning the search space $A_\ell$, thus solving fewer subproblems, relying on a bounding technique. For each of the leader's actions, we compute the utility he would obtain if the followers played a CE. This yields a UB, as the set of correlated strategies is a (strict) superset of that of mixed strategies. We can thus iterate over $ i \in A_\ell$ and solve one of our formulations with $\delta = e_i$ only if the UB with $\delta = e_i$ is better than the best solution found thus far.
The algorithm reads:
\begin{algorithmic}[1]
\begin{scriptsize}
\FOR {$i \in A_\ell$}
\STATE $UB(i) = BestCorrelatedEquilibrium(i)$
\ENDFOR
\STATE $A_\ell = DescendingSort (A_\ell, UB)$
\STATE $LB = -\infty$
\FOR{$i \in A_\ell$ and $UB(i)>LB$}
\STATE $LB = \max\{LB, Utility(e_i)\}$
\ENDFOR
\end{scriptsize}
\end{algorithmic}
$BestCorrelatedEquilibrium(i)$ finds a UB with $\delta = e_i$ by computing a CE in polynomial time via linear programming, along the lines of~\cite{shoham-book}. After sorting the leader's actions in decreasing order of UB via $DescendingSort (A_\ell, UB)$, the algorithm iterates over $A_\ell$, computing with $Utility(e_i)$ the exact utility when $\delta = e_i$,
only if $UB(i)$ is sufficiently promising. In our implementation, $Utility(e_i)$ solves the same oracle formulations adopted in the black box method.


\subsubsection{O--NF--LPFM--III} For $\delta \in \{0,1\}^{m_\ell}$, the (quadratic) Constraints~\eqref{zik} in O--NF--LMFM--III can be dropped in favor of the following three linear constraints:

\vspace{-.5cm}
\begin{scriptsize}
\begin{align}
y^{ij}_f \leq \delta^i & \quad \forall i \in A_\ell, f \in F, j \in A_f \label{16-1}\\
y^{ij}_f \leq \rho^j_f & \quad \forall i \in A_\ell, f \in F, j \in A_f\label{16-2}\\
y^{ij}_f \geq \delta^i + \rho^j_f - 1 &  \quad \forall i \in A_\ell, f \in F, j \in A_f\label{16-3}.
\end{align}
\end{scriptsize}
\vspace{-0.5cm}

\noindent Together with $y^{ij}_f \geq 0$, these constraints constitute the so--called McCormick envelope~\cite{mccormick1976computability} of the set $\{ (y^{ij}_f, \delta^i, \rho^j_f) \in [0,1]^3: z^{ij}_f = \delta^i \rho^j_f\}$. When either $\delta^i \in \{0,1\}$ or $\rho^j_f \in \{0,1\}$, the envelope yields an exact reformulation. Thus, the only nonlinear constraints in O--NF--LPFM--III are Constraints~\eqref{zijk}. The resulting formulation is the following one:

\begin{scriptsize}
\begin{align}
\max \quad \sum_{i \in A_\ell}\sum_{j \in A_1}\sum_{k \in A_2} z^{ijk}{U}_{\ell}^{ijk} & \quad \text{s.t.} \label{objectino}\\
u_f^j = \sum_{i \in A_\ell} \sum_{k \in A_{f'}}  y_{f'}^{ik}  {U}_f^{ijk} & \quad \forall f\in F, j \in A_f \label{cons22}\\
\sum_{i \in A_\ell}\delta^i = 1, \delta \in\{0,1\}^{m_\ell}\\
\textnormal{Constraint~\eqref{rho}, \eqref{cons1}--\eqref{consSupp}, \eqref{zijk}--\eqref{cons8}, \eqref{16-1}--\eqref{16-3}.} \label{finalino}
\end{align}
\end{scriptsize}

\subsubsection{O--PM--LPFM--III} In O--PM--LMFM--III, the only nonlinearities are due to the constraints $y^{ij}_f = \delta^i \rho^j_f$. Applying, due to $\delta \in \{0,1\}^{m_\ell}$, the McCormick envelope, we can remove all the nonlinearities from the problem. O--PM--LPFM--III is, thus, a mixed-integer linear program and it reads as follows:

\begin{scriptsize}
\begin{align}
\hspace{-0.5cm}\max \quad \sum_{i \in A_\ell} \sum_{f \in F}\sum_{j \in A_f} y_{f}^{ij}  U_{\ell f}^{ij} & \quad\text{s.t.} \\
{u}_f^j = \sum_{i \in A_\ell} \delta^i U_{f\ell}^{ij} + \sum_{k \in A_{f'}} \rho^{k}_{f'} U_{ff'}^{jk} & \quad \forall f \in F, j \in A_f\\
\sum_{i \in A_\ell}\delta^i = 1, \delta \in\{0,1\}^{m_\ell}\\
\textnormal{Constraints~\eqref{rho}, \eqref{cons1}--\eqref{consSupp},\eqref{cons5}, \eqref{cons7}, \eqref{16-1}--\eqref{16-3}.}
\end{align}
\end{scriptsize}

\section{Experimental evaluation}\label{sec:experimental}

Our testbed is composed of instances of two GAMUT classes, {\tt (Uniform) RandomGames} (which are normal form games) and {\tt PolymatrixGames}, generated with payoffs in $[0,100]$ and the same number of actions $m$ for each agent, with 10 different instances per value of $m$ and game class.
For some experiments, we will also consider other GAMUT classes of structured normal form games, as better explained in the following.
%
We experiment on games with an increasing $m$, so to assess how our methods scale with the game size. We select $m \in \{2,3,\ldots,10,15,\ldots, 50\}$ for $n=3$ (2 followers) and $m \in \{2,3,\ldots,10\}$ for $n \geq 4$ ($\geq 3$ followers).
We will compare the results of our experiments w.r.t. \emph{computing time} (in seconds) and (multiplicative) \emph{optimality gap}\footnote{\scriptsize The optimality gap is defined as $\min\{\frac{\textnormal{UB}-\textnormal{LB}}{\textnormal{LB}} \, 100, 10^5\}\%$, where LB and UB are, resp., the largest lower bound (corresponding to the best feasible solution) and the smallest upper bound found by the solver within the time limit. The $\min$ operator prevents an unbounded value for LB = 0. Thus, an optimality gap of $10^{5}$ highlights that the method fails to produce a useful solution as, due to the payoffs being in $[0,100]$, {\em any} strategy of the leader can achieve, at least, a utility of 0.}. For both values, we will report the arithmetic average for each game class and value of $m$ over the 10 corresponding instances.

We adopt five solvers: BARON 13.0.1 and SCIP 3.0.0 (for globally optimal solutions to every formulation, apart from O--PM--LPFM--III, which is an MILP),  CPLEX 12.6.2  (for globally optimal solutions to O--PM--LPFM--III, as well as to the oracle formulation' for PM games in the implicit enumeration and black box methods), SNOPT 7.4.2 (for locally optimal solutions to the formulations with purely continuous variables),
and RBFOpt 1.1.0 as a heuristic for both the optimistic and pessimistic cases of LFE--N. 
The O--NF--LPFM--Implicit--Enumeration algorithm is implemented in C. The experiments are run on a UNIX computer with a dual quad--core CPU at 2.33 GHz, equipped with 8~GB of RAM, within a time limit of 3600 seconds. Each algorithm is run using a single thread. For the exact methods, we halt the execution whenever the optimality gap reaches\footnote{\scriptsize Preliminary experiments with four tolerance values, namely, $10^{-12}\%$, $10^{-9}\%$, $10^{-6}\%$, and $10^{-3}\%$, showed, for a larger tolerance, a negligible reduction in computing time by, at most and only in few instances, $2.5\%$ with SCIP and $7.0\%$ with BARON. The stricter tolerance was thus preferred.} $10^{-12}\%$.

\subsection{O--NF--LMFM--I, II, and III ($n = 3$)} 
\begin{figure}[!t]
\subfigure[Average times (BARON )]{\includegraphics[width=0.23\textwidth]{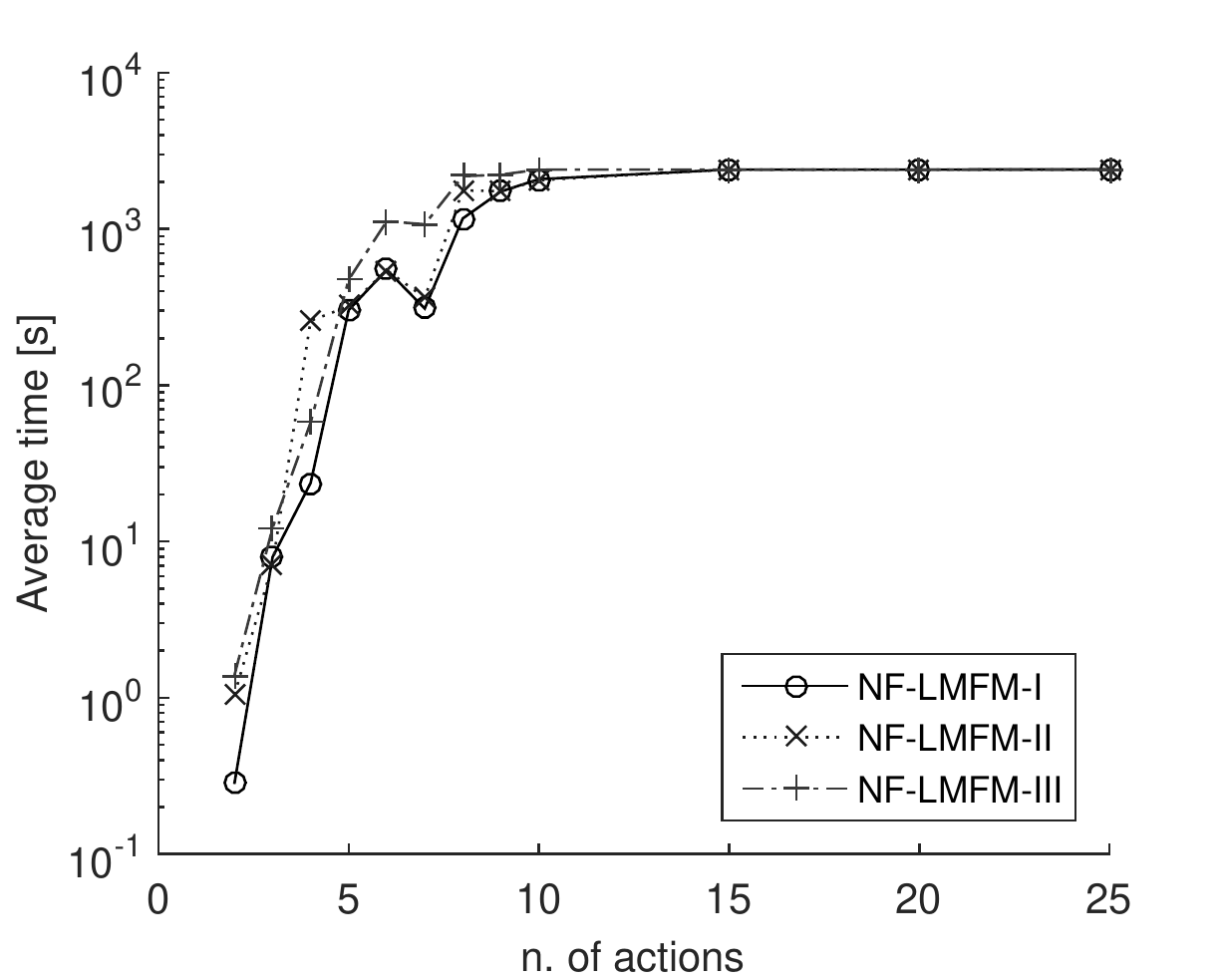}}
\subfigure[Average gaps (BARON)]{\includegraphics[width=0.23\textwidth]{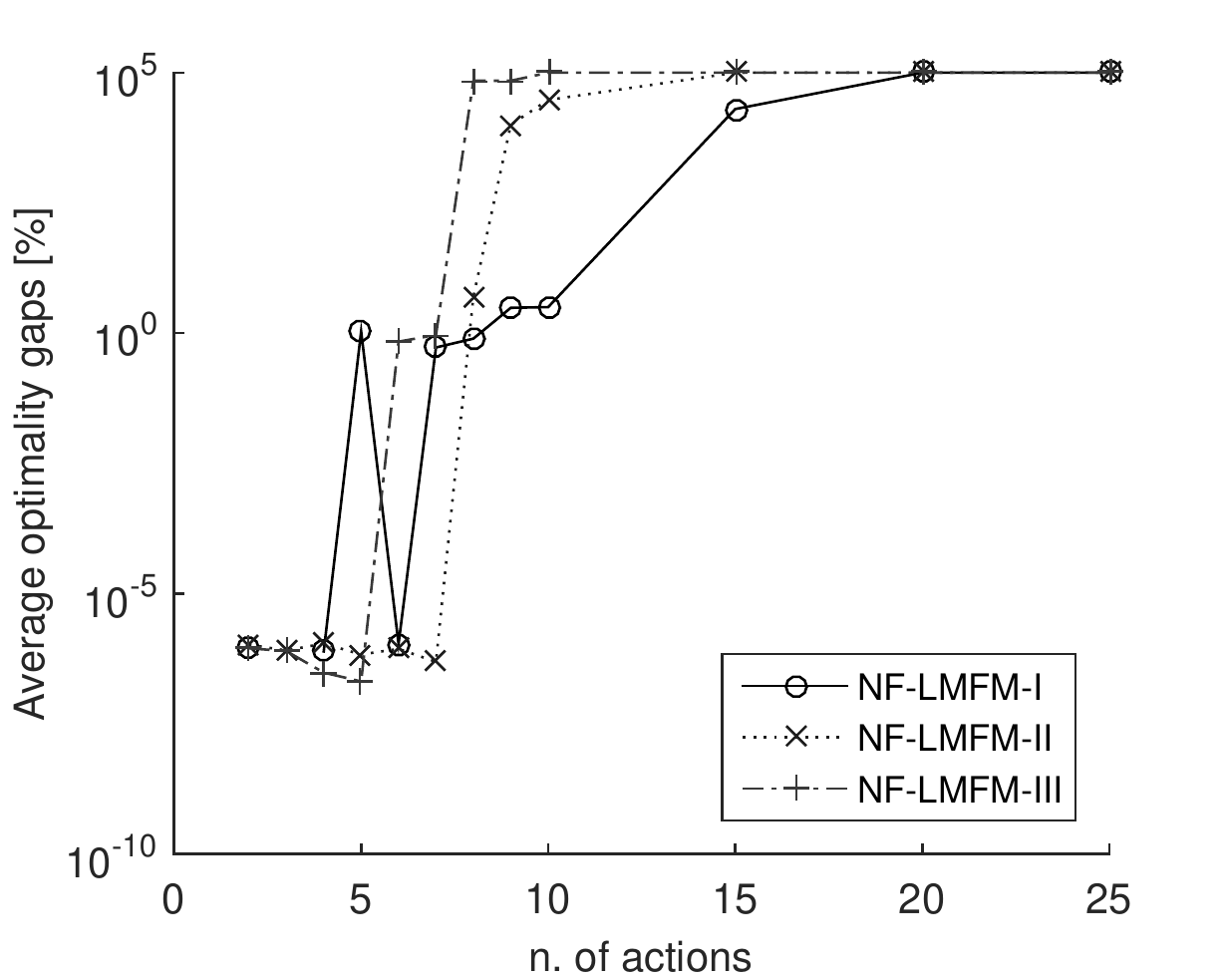}}\\
\subfigure[Average times (SCIP)]{\includegraphics[width=0.23\textwidth]{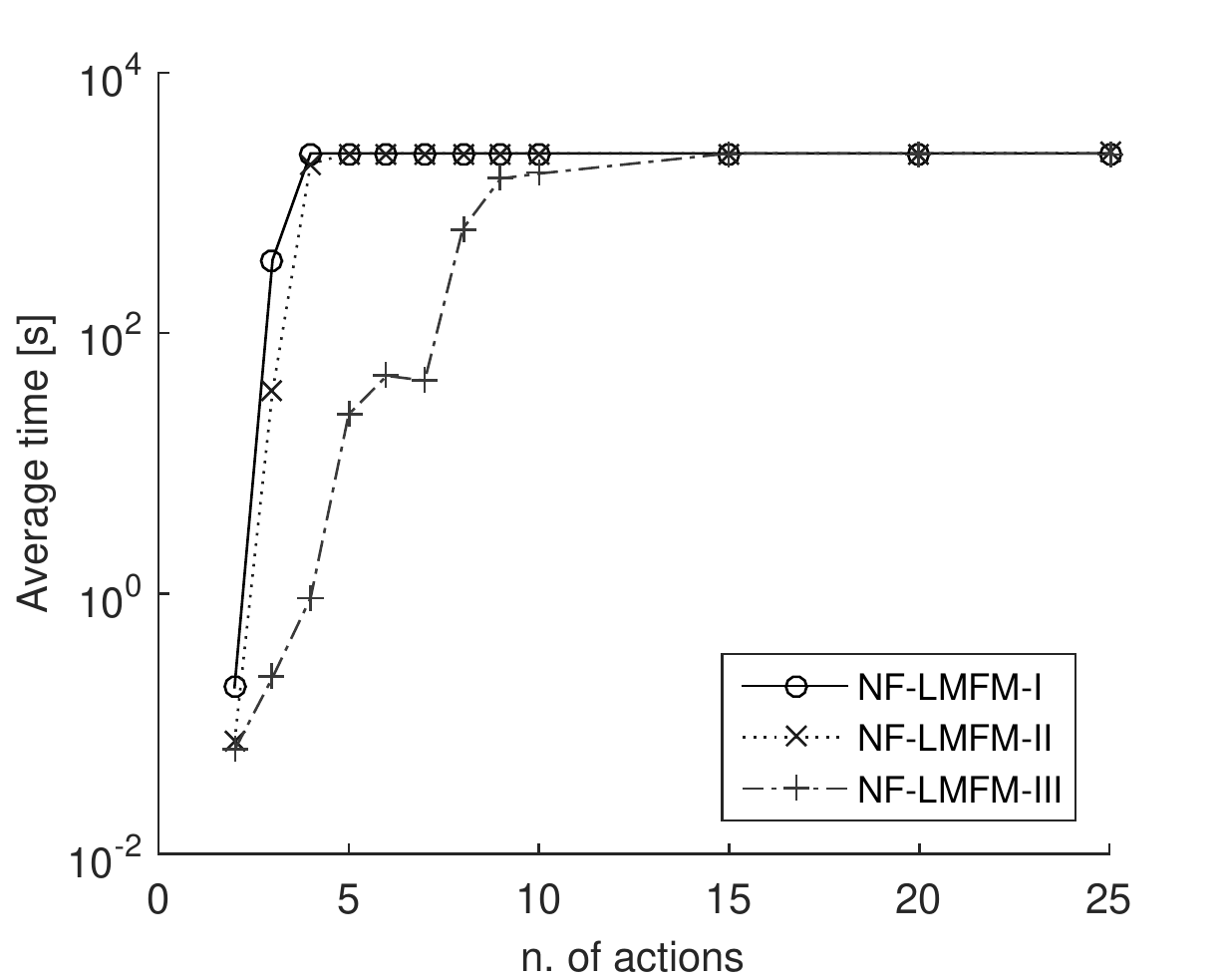}}
\subfigure[Average gaps (SCIP)]{\includegraphics[width=0.23\textwidth]{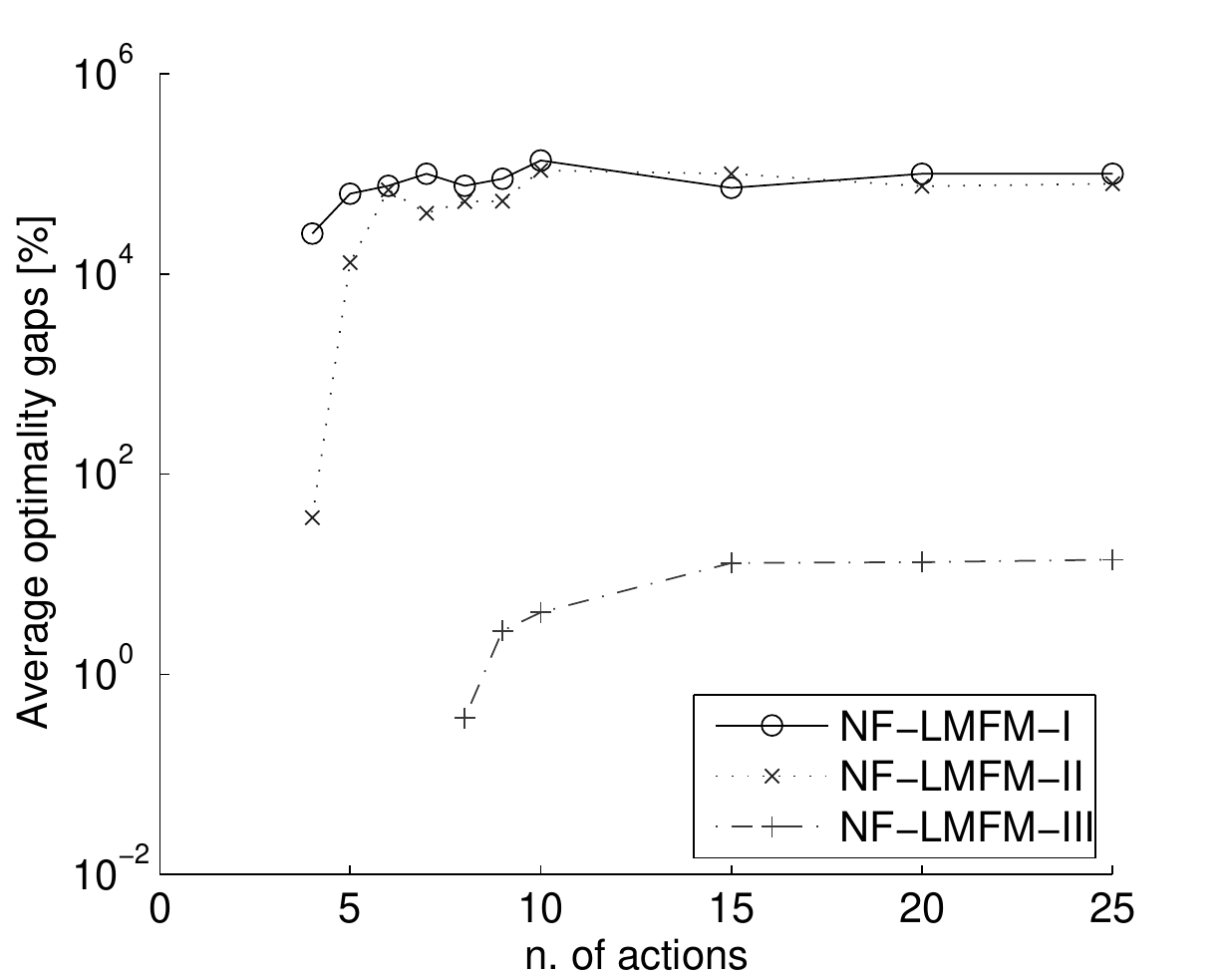}}
\caption{Computing times and optimality gaps for NF--LMFM formulations.}
\label{fig:baron_scip_nf}
\end{figure}

We compare the different NF formulations when solved with BARON and SCIP. The average computing time and optimality gap for each combination of formulation and solver is reported, for {\tt RandomGames} instances, in Fig.~\ref{fig:baron_scip_nf}, as a function of $m$. (For the sake of clarity we report data for $m$ up to 25.)

The results are opposite for the two solvers. BARON better performs on O--NF--LMFM--I (the purely continuous formulation), while SCIP better performs on O--NF--LMFM--III (the ``reformulated'' formulation obtained after removing nonquadratic  terms from O--NF--LMFM--II, containing binary variables and extra valid constraints). These results suggest O--NF--LMFM--I with BARON and O--NF--LMFM--III with SCIP as the formulations which are most efficiently solved with each global solver. 

Inspecting Fig.~\ref{fig:baron_scip_nf}, we notice that, with SCIP, O--NF--LMFM--III always outperforms O--NF--LMFM--II, showing that the solver is incapable of automatically constructing the reformulation obtained with O--NF--LMFM--III.

As to the computing times, the largest $m$ for which at least a game is solved to optimality by BARON within the time limit is $m = 8$ for O--NF--LMFM--I and $m = 7$ for the other formulations. With SCIP, we reach $m = 9$ with O--NF--LMFM--III, and $m = 3$ with the other ones. Although SCIP with O--NF--LMFM--III and BARON with O--NF--LMFM--I have a similar performance, the former requires a shorter time than the latter for any number of actions.

\begin{figure}[!t]
\subfigure[O--NF--LMFM--I (BARON)]{\includegraphics[width=0.23\textwidth]{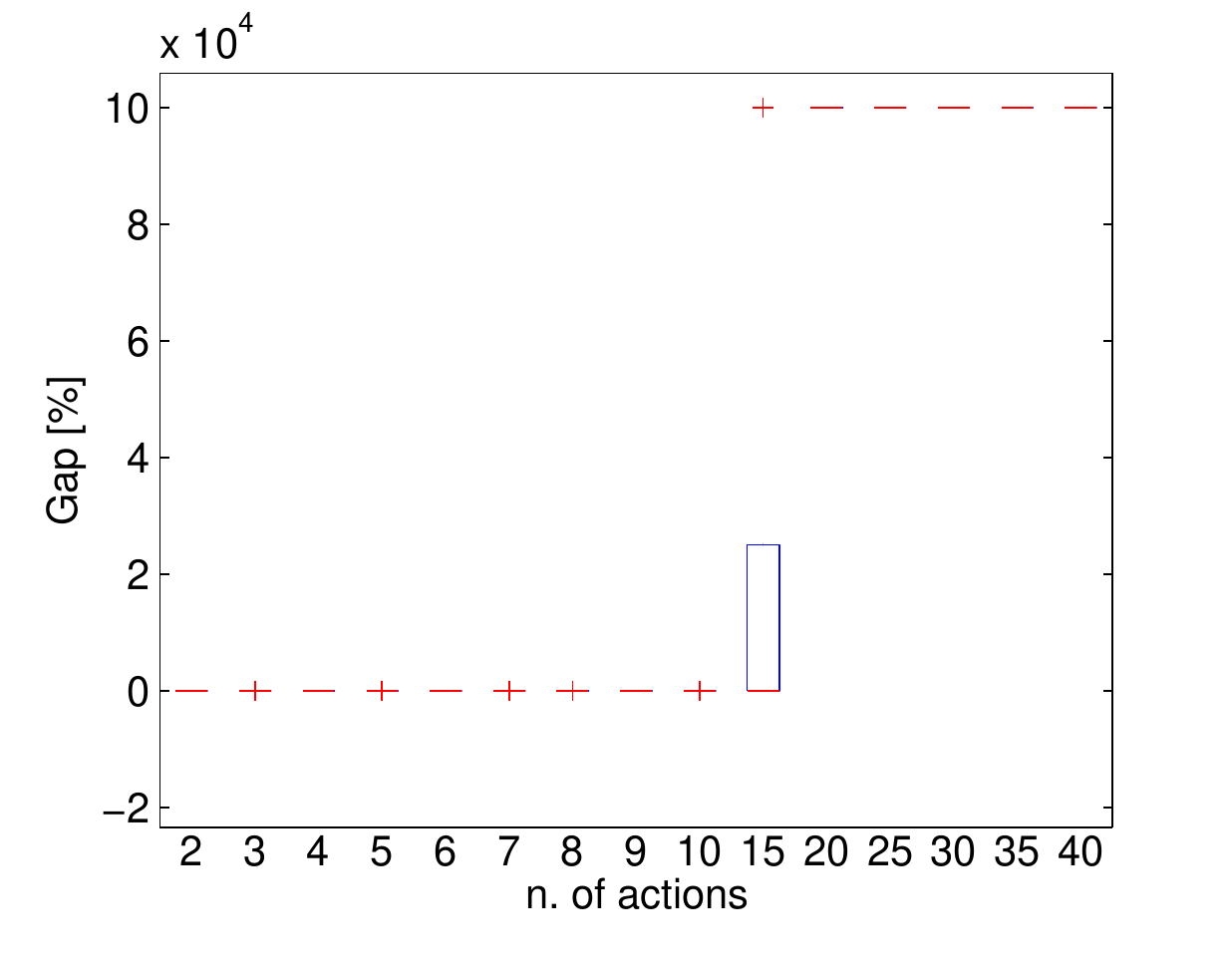}}
\subfigure[O--NF--LMFM--III (SCIP)]{\includegraphics[width=0.23\textwidth]{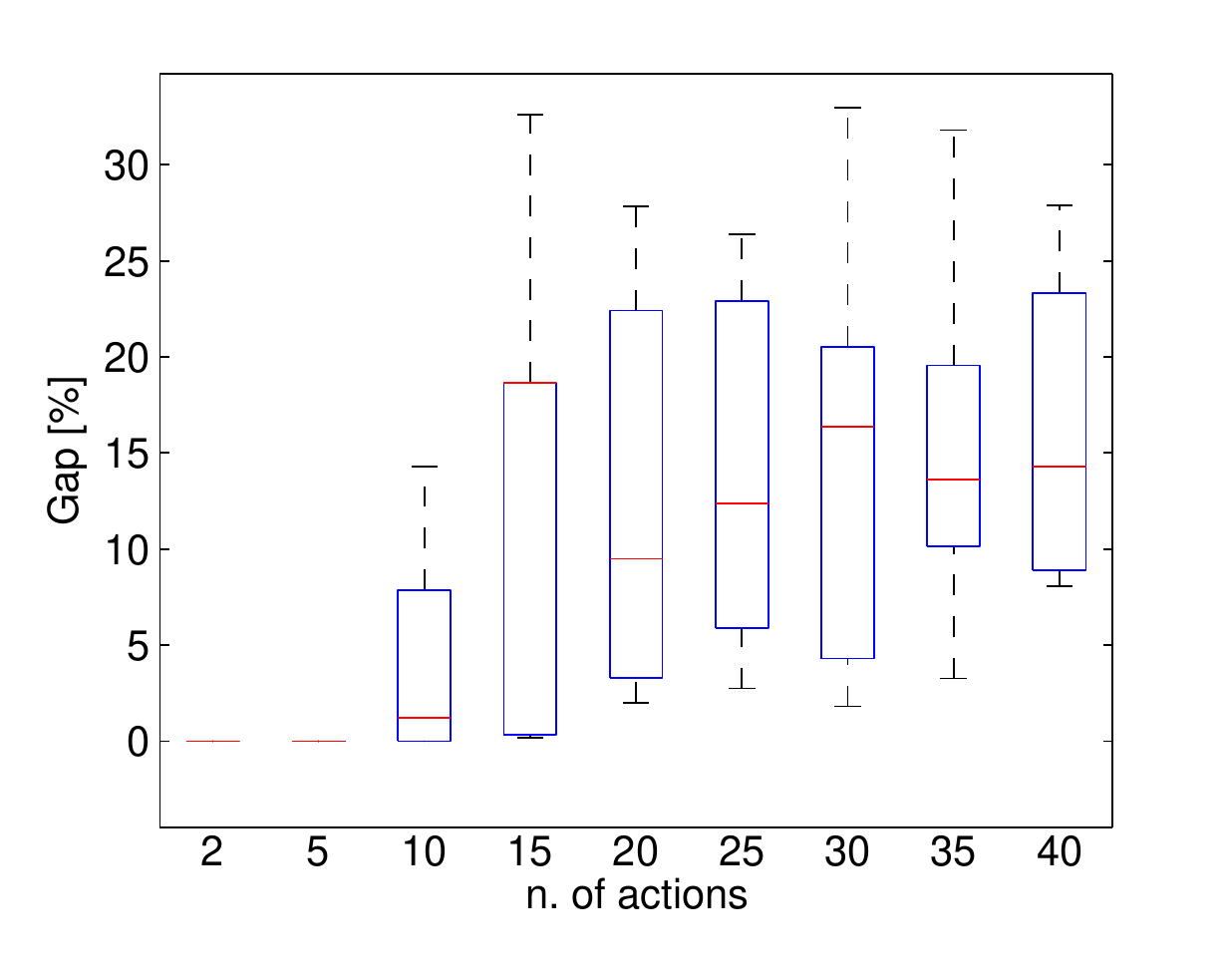}}
\caption{Detailed results on optimality gaps with the most efficient solver/formulation pairs for NF games.}
\label{fig:gaps_boxplots}
\end{figure}

\begin{figure}[!b]
\subfigure[Multiplicative gaps]{\includegraphics[width=0.23\textwidth]{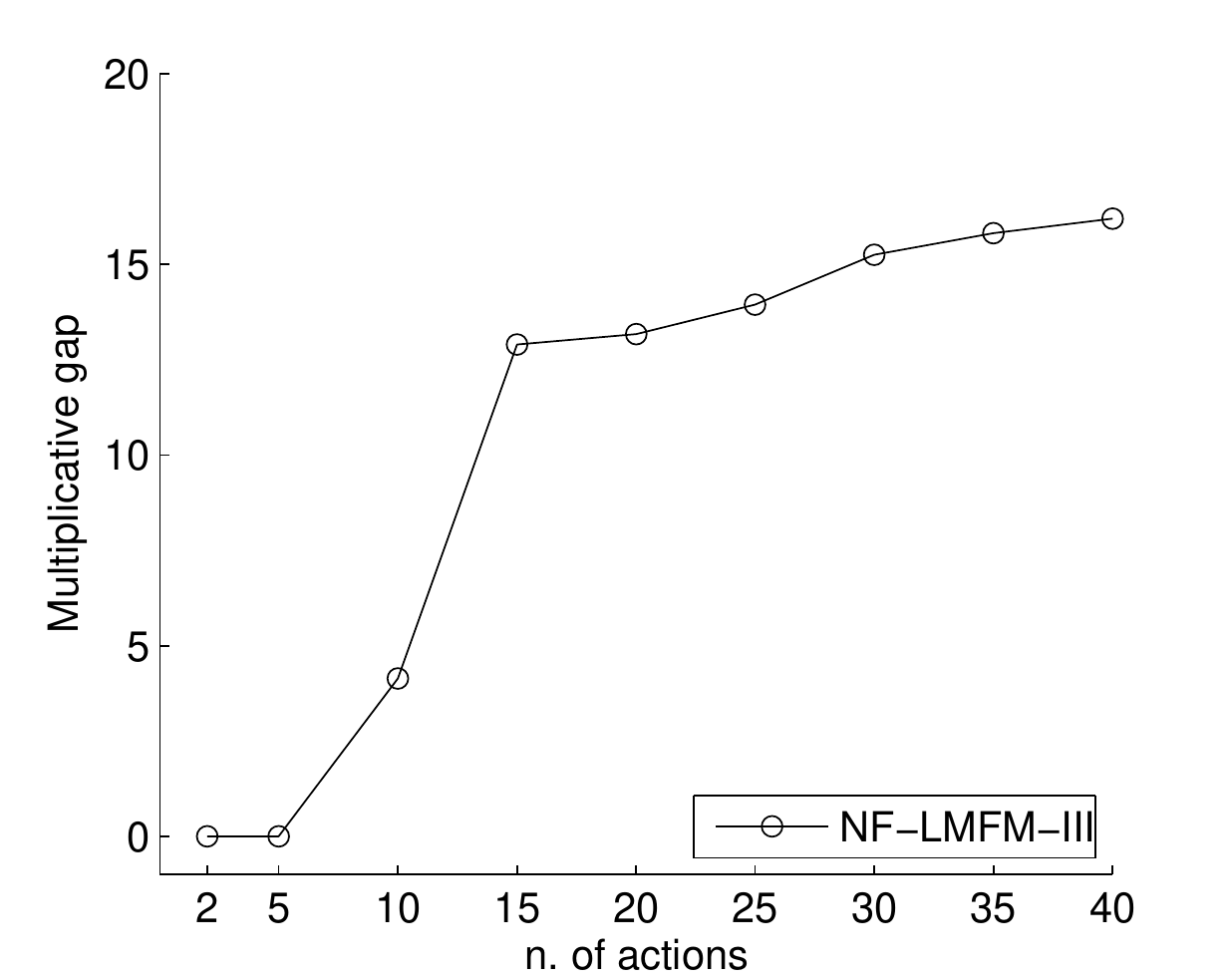}}
\subfigure[Additive gaps]{\includegraphics[width=0.23\textwidth]{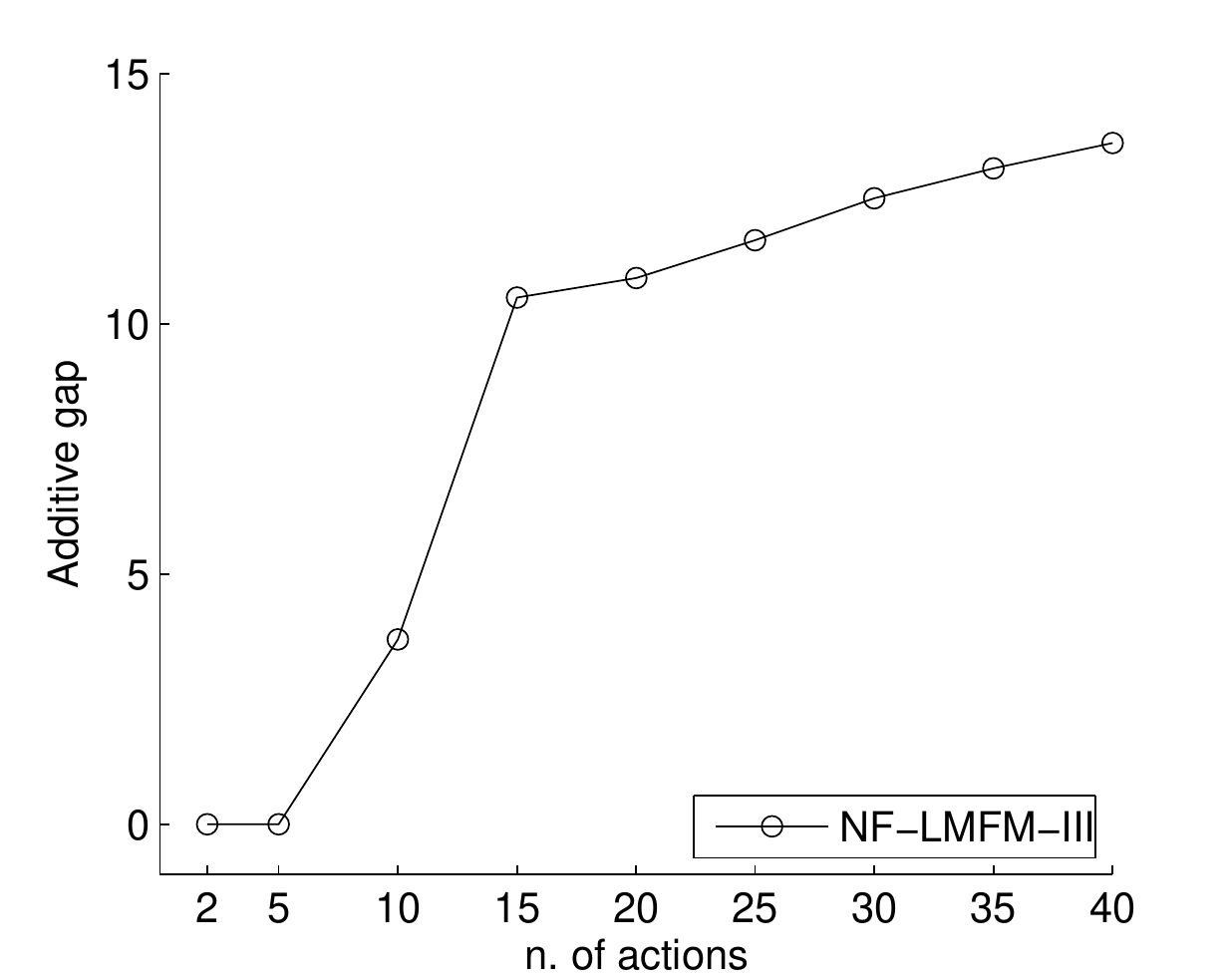}}
\caption{Optimality gaps with SCIP with NF--LMFM--III.}
\label{fig:mutliplicativevsadditive}
\end{figure}

We report in Fig.~\ref{fig:gaps_boxplots} the corresponding statistical distribution of average gap trends up to $m=40$ with the two more efficient solver/formulation pairs.\footnote{\scriptsize In this and subsequent boxplots, the red dash indicates the median, the box extends from the 25th to the 75th percentile while dotted lines denote the whole sample distribution; outliers are marked with a red mark.}Also, in Fig.~\ref{fig:mutliplicativevsadditive}, we report multiplicative and additive gaps for SCIP with O--NF--LMFM--III (the most efficient profile).
When considering such obtained optimality gaps, SCIP remarkably outperforms BARON. As can be seen in Figure~\ref{fig:baron_scip_nf}~(b),(d), the gap achieved by BARON with O--NF--LMFM--I reaches $10^{5}$\% for $m \geq 20$ (where the LB is 0 for almost all the games, as a consequence of the solver failing to find a feasible solution in the time limit), while the average optimality gap achieved by SCIP with NF--LMFM--III, see Fig.~\ref{fig:mutliplicativevsadditive}(a), is always smaller than 30\% even for $m = 40$ while the worst case gap can be upper bounded by 35\% (see Fig.~\ref{fig:gaps_boxplots}). Surprisingly, such result shows how we achieved an almost constant empirical approximation factor contrarily to what the intrinsic difficulty of the problem would suggest, namely an exponential quality degradation as the number of actions grows. Moreover, these results show that SCIP with O--NF--LMFM--III always finds a feasible solution (a Nash equilibrium) for the followers' game, differently from the other pairs of solver and formulation.
For completeness, for the same pair of solver and formulation, we also measured the additive optimality gap, see Fig.~\ref{fig:mutliplicativevsadditive}(b), which, as of our experiments, is $\leq 15$ for up to $m = 40$\footnote{\scriptsize Additive optimality gap is defined as UB$-$LB.}. Games with $m\geq 45$ actions were not solved due to memory limits. Thanks to its very low optimality gaps (both multiplicative and additive), SCIP with O--NF--LMFM--III may constitute a valid (empirical) approximation algorithm yielding, for up to $m = 40$, a $\frac{6.5}{10}$--approximation\footnote{\scriptsize Use $\frac{\textnormal{OPT}-\textnormal{LB}}{\textnormal{OPT}} \leq \frac{\textnormal{UB}-\textnormal{LB}}{\textnormal{LB}} \leq \frac{3.5}{10}=35\%$.}.

This trend is substantially confirmed by results we obtained with other GAMUT normal--form game classes. In such experiments, we evaluate the aforementioned formulation/solver pairs for NF games of the eight GAMUT classes which we report, for the sake of clarity, in Table~\ref{table:classes}. In detail: for each these classes we solve 10 random instances with 2 followers and $m=8$ actions per player; for each game instance, we compute an optimistic LFN--E by solving formulations O--NF--LMFM--I and O--NF--LMFM--III with BARON and SCIP, respectively.

\begin{figure}[!htbp]
\includegraphics[width=0.45\textwidth]{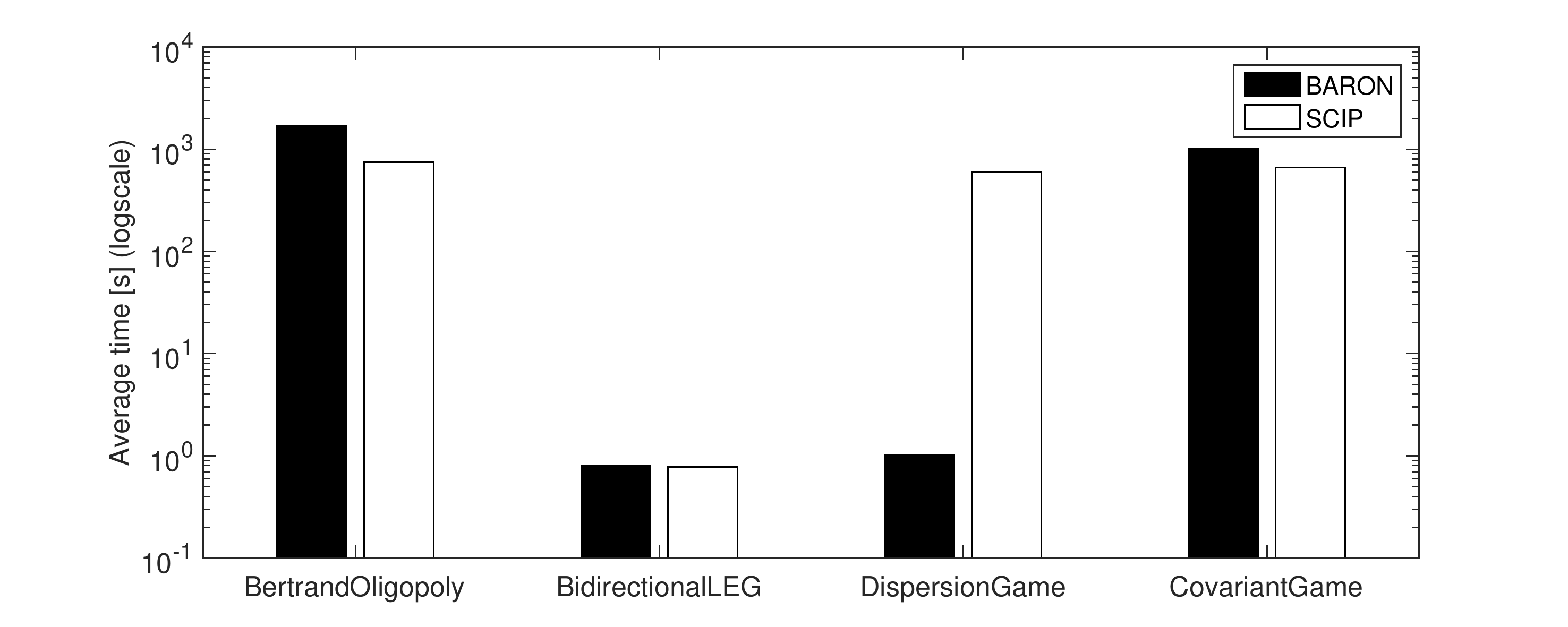}
\includegraphics[width=0.45\textwidth]{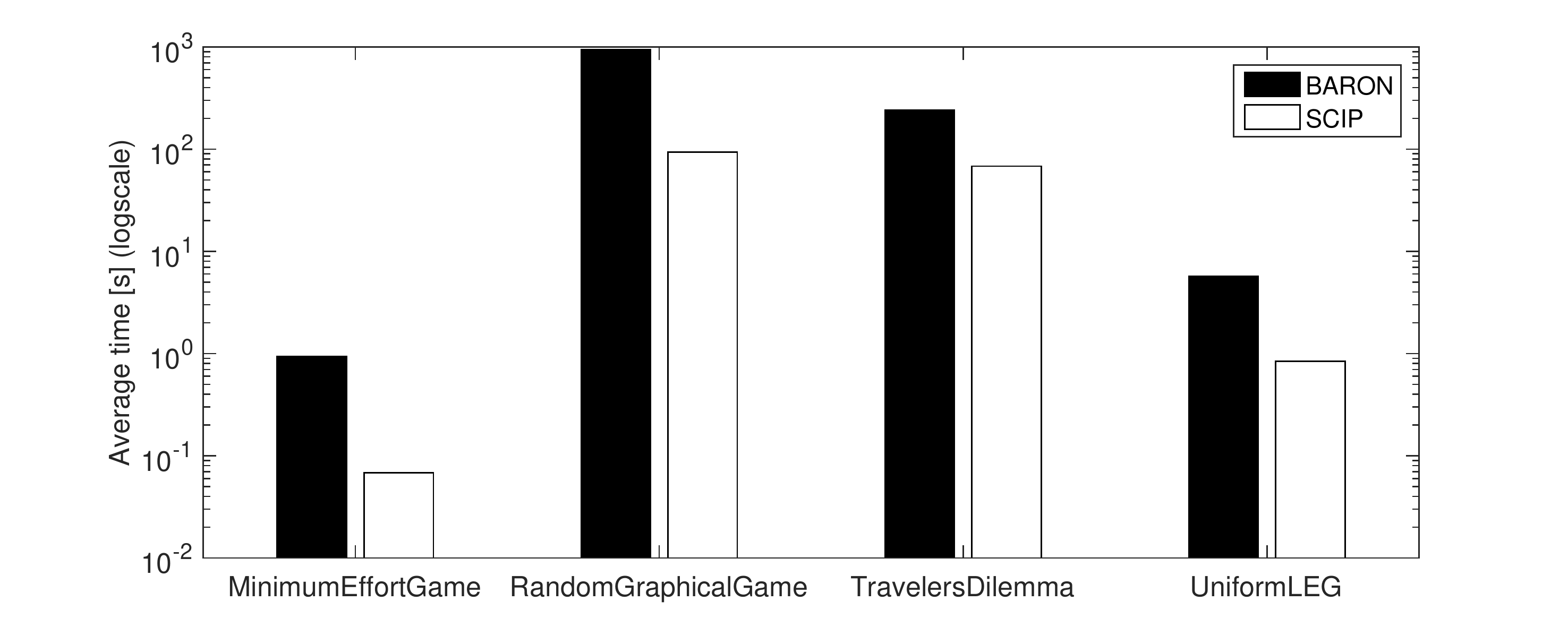}
\caption{Computing times obtained when solving formulation O--NF-LMFM-I with BARON and formulation O--NF--LMFM-III with SCIP for different GAMUT classes of games.}
\label{fig:other_classes}
\end{figure}

\begin{table}[!h]
\caption{Additional GAMUT game classes}\label{table:classes}
\begin{tabular}{|c|c|}
\hline
{\tt BertrandOligopoly} & {\tt BidirectionalLEGs}\\
\hline
{\tt MinimumEffortGames} & {\tt RandomGraphicalGames}\\
\hline
 {\tt DispersionGames} & {\tt CovariantGames}\\
\hline
 {\tt TravelersDilemma} & {\tt UniformLEGs}\\
\hline
\end{tabular}
\end{table}
The average computing times reported in Fig.~\ref{fig:other_classes} substantially confirm the trends we observed for {\tt RandomGames}, with SCIP outperforming BARON, on average, most of the times. Surprisingly, this trend becomes radically different for {\tt DispersionGames}, where SCIP performs less efficiently than for the other classes of games, achieving computing times which are considerably larger than those obtained with BARON. This is due to SCIP failing to solve two game instances within the time limit.

\subsection{O--PM--LMFM--I, II, and III ($n = 3$)} 
\begin{figure}[!b]
\subfigure[Average times (SCIP)]{\includegraphics[width=0.23\textwidth]{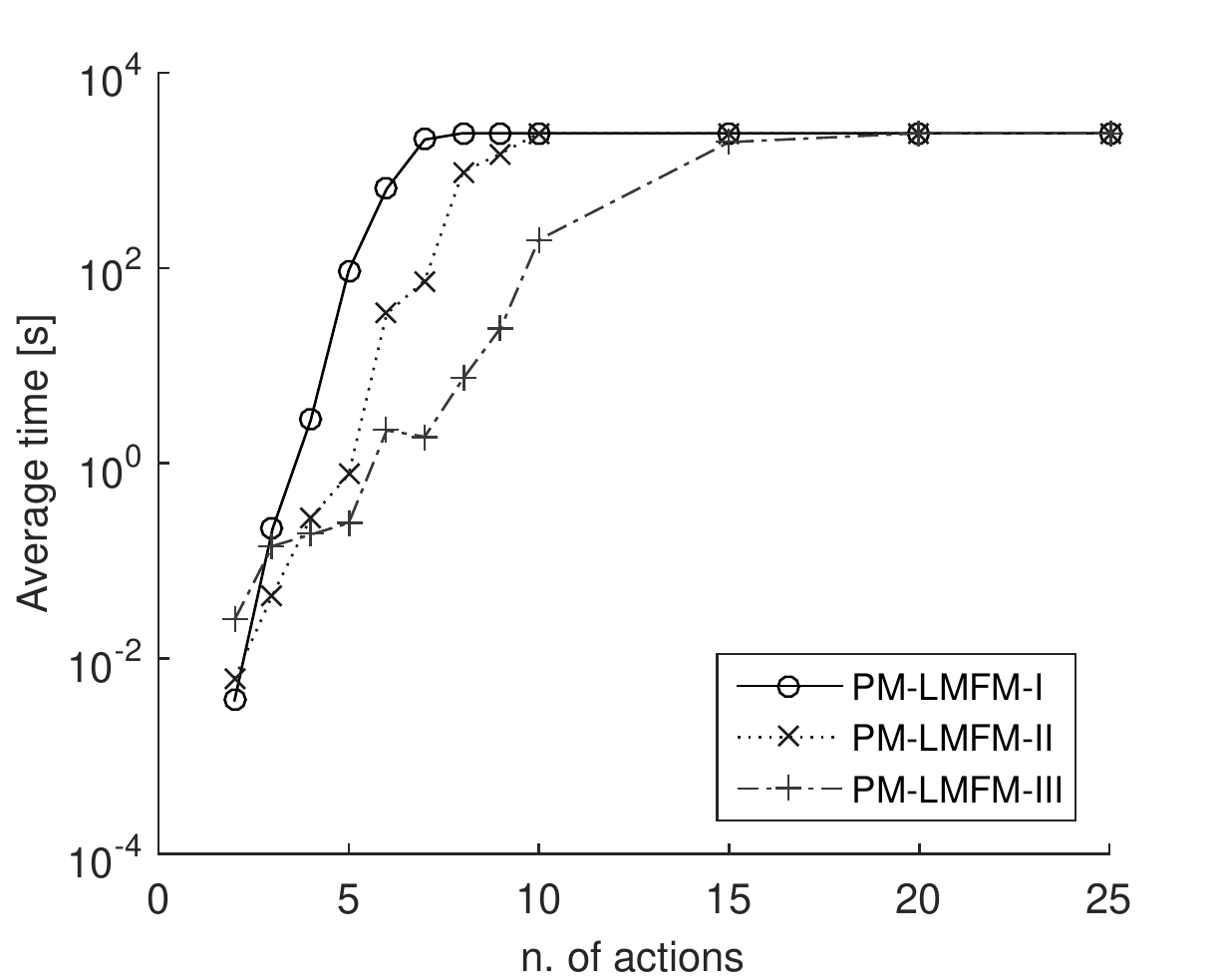}}
\subfigure[Average gaps (SCIP)]{\includegraphics[width=0.23\textwidth]{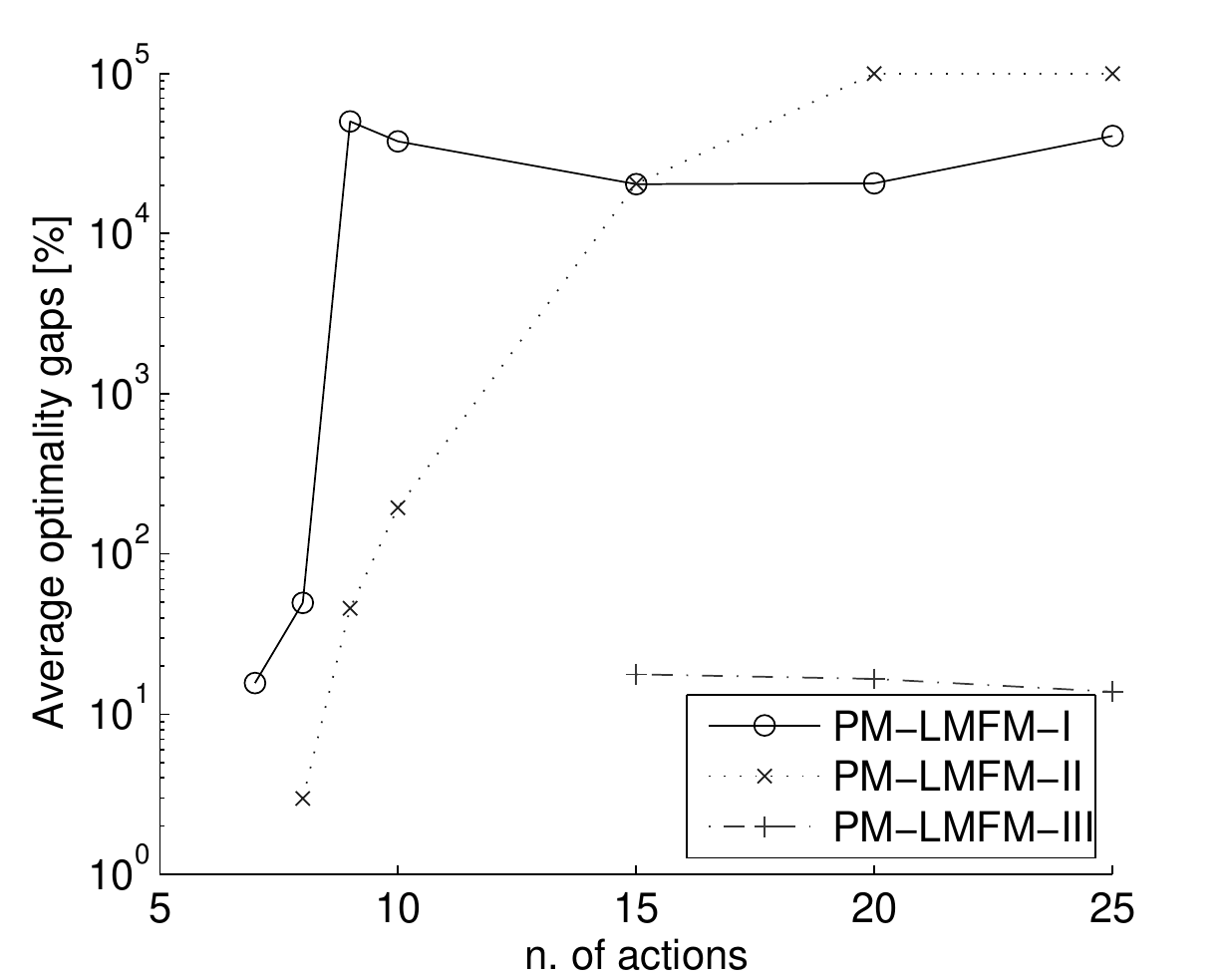}}
\caption{Computing times and optimality gaps with SCIP with PM--LMFM formulations.}
\label{fig:polymatrix_scip}
\end{figure}

\begin{figure}[!t]
\subfigure[O--PM--LMFM--III (SCIP)]{\includegraphics[width=0.23\textwidth]{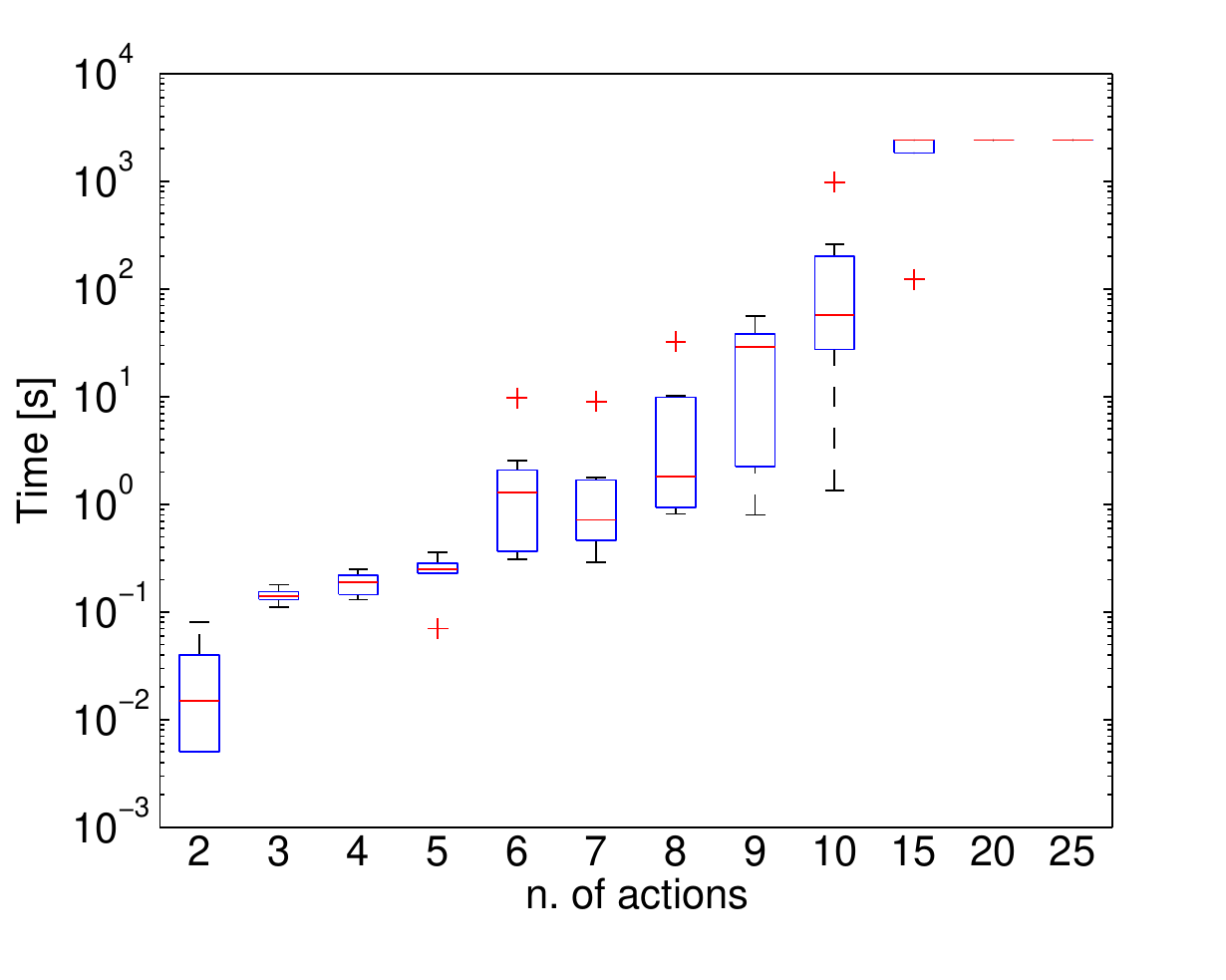}}
\subfigure[O--PM--LMFM--III (SCIP)]{\includegraphics[width=0.23\textwidth]{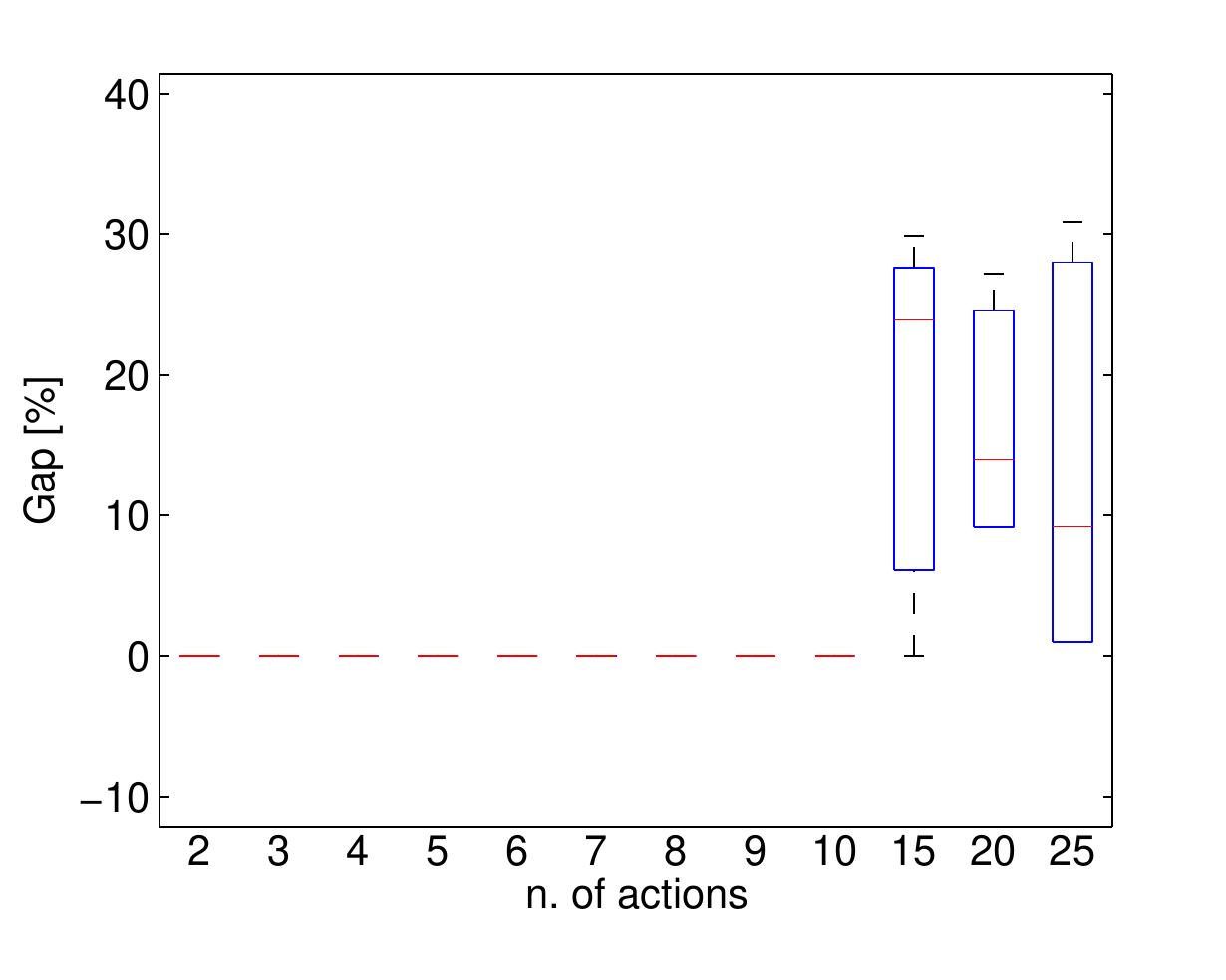}}
\caption{Detailed results on computing time and optimality gaps with the most efficient solver/formulation pair for PM games.}
\label{fig:times_gap_PM_boxplot}
\end{figure}

In Fig.~\ref{fig:polymatrix_scip}, we report the computing times and the optimality gaps on {\tt PolymatrixGames} obtained with SCIP (the results obtained with BARON are omitted for reasons of space). Within the time limit, the largest $m$ for which at least an instance is solved to optimality is $m=15$, while, for $m \le 10$ all instances are solved to optimality. Although the complete results are not shown for reasons of space, for this class of games we can handle instances with up to $m=50$ before SCIP runs out of memory. We register a  worst case optimality gap below $30\%$ (of the same order as the optimality gap obtained with NF games) for up to $m = 50$. This suggests that, on PM games, SCIP with O--PM--LMFM--III may constitute an empirical approximation algorithm for instances with up to $m = 50$, empirically yielding a (as in the normal--form case, almost constant) $\frac{7}{10}$--approximation. Similarly to what done before, we report (see Fig.~\ref{fig:times_gap_PM_boxplot}) statistical details of the results obtained with such formuation for time and optimality gaps.

\subsection{O--NF--LMFM--I, local optimization ($n = 3$)}  
\begin{figure}[!t]
\subfigure[Average times (SNOPT)]{\includegraphics[width=0.23\textwidth]{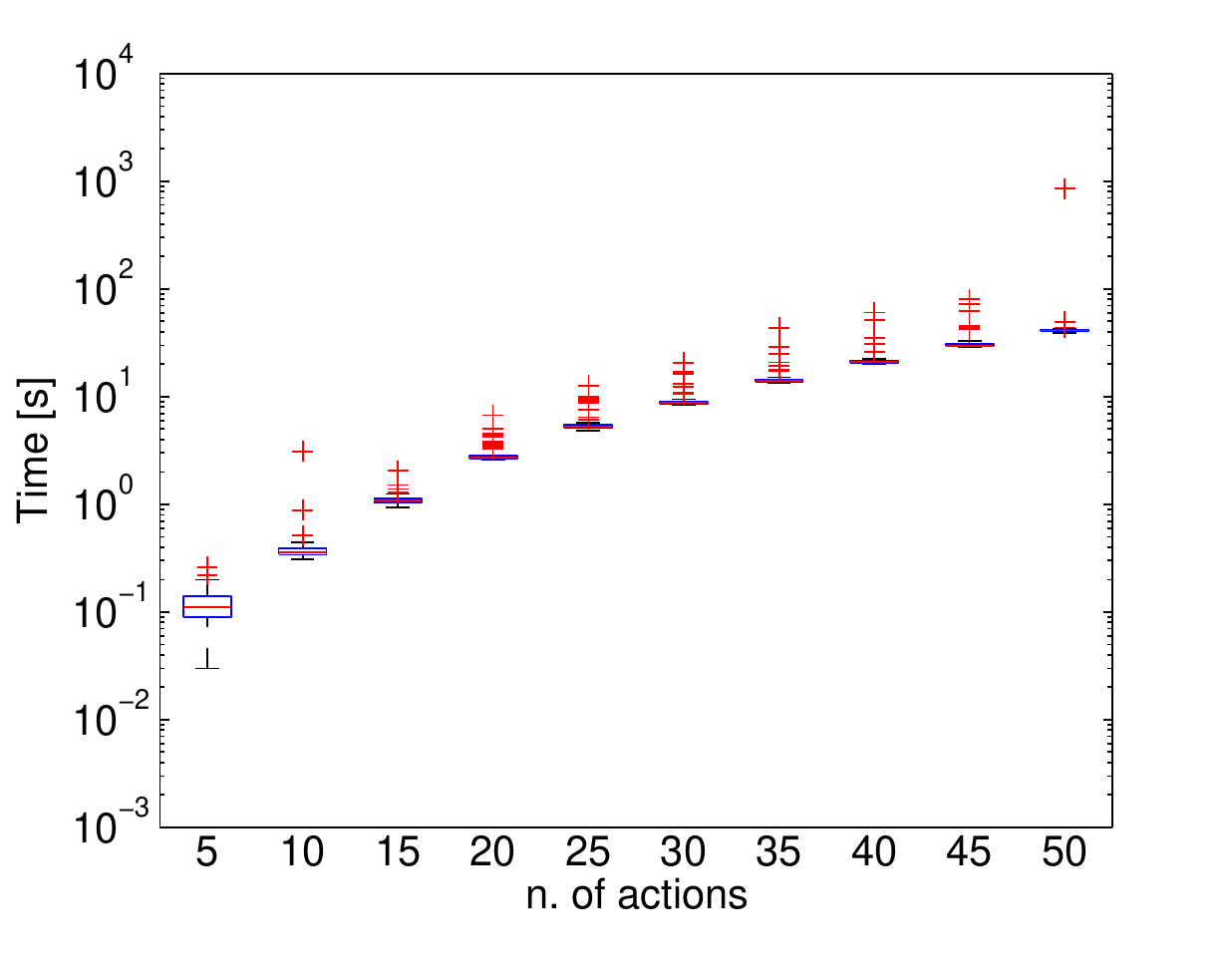}}
\subfigure[Average gaps (SNOPT)]{\includegraphics[width=0.23\textwidth]{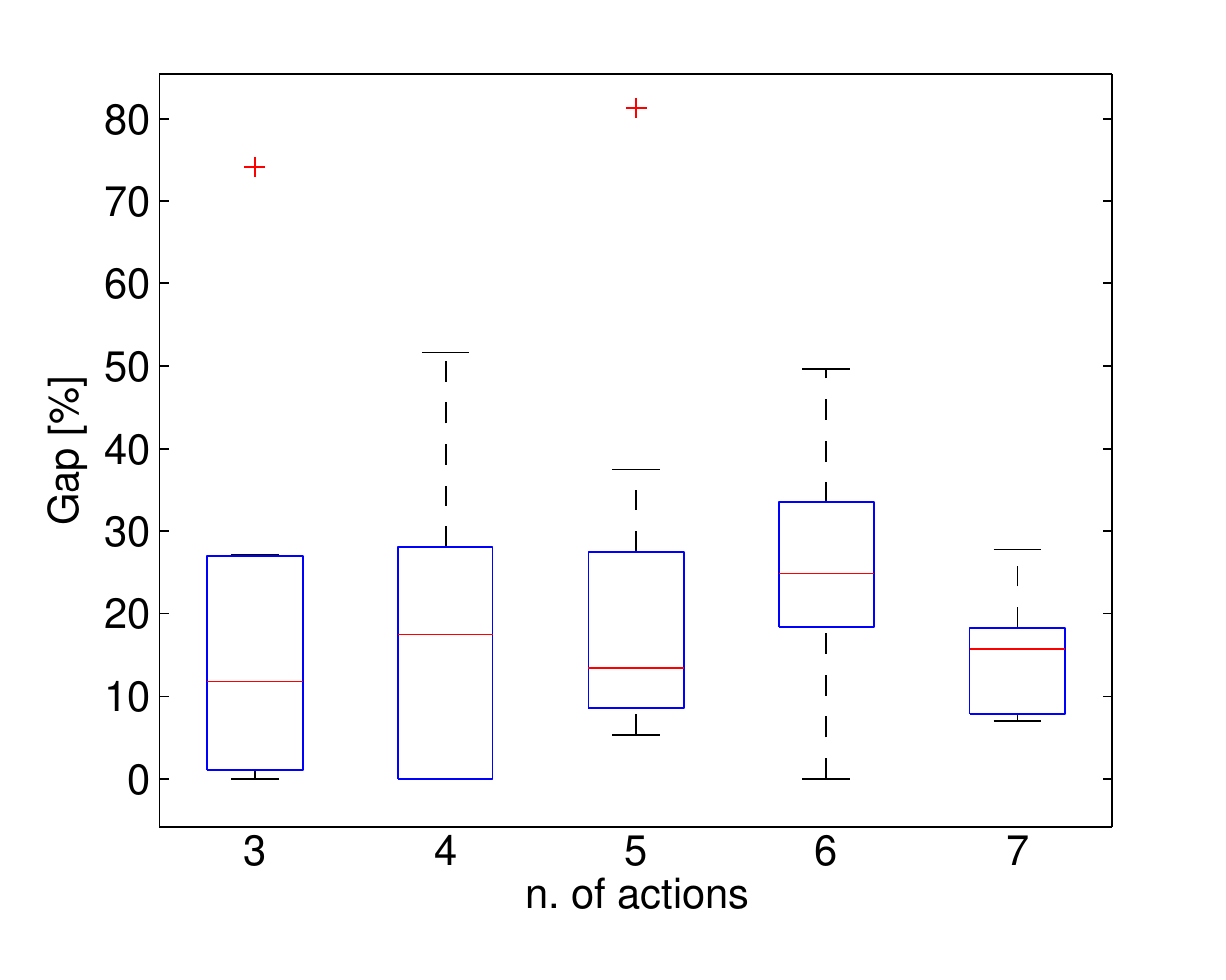}}
\caption{Computing times and $\frac{\textnormal{LB}}{\textnormal{OPT}}$ ratios obtained with SNOPT with O--NF--LMFM--I within 30 random restarts.}
\label{fig:snopt}
\end{figure}

We report, in Fig.~\ref{fig:snopt}, the experimental results obtained with SNOPT for {\tt RandomGames}. Due to the local optimization nature of the solver for nonconvex problems, to obtain statistically more relevant results, we run 30 restarts with different initial starting solutions, sampled uniformly at random from the simplices of the strategies of the three agents, and return the best solution found. Fig.~\ref{fig:snopt}(a) shows that the computing times with SNOPT (cumulated over the 30 random restarts) are much shorter than the computing times required by the global solvers, allowing for the solution (to a local optimum) of almost all the instances with $m = 50$ within the time limit. Differently, as shown in Fig.~\ref{fig:snopt}(b), the quality of the solutions returned by SNOPT, measured as their ratio over the value of an optimal solution, as found by SCIP or BARON, is rather poor even with very few actions. Indeed, the median of the ratios is between $10\%$ and $20\%$ for games with up to $m=7$. This suggests that resorting to local optimization (with random restarts) can be effective only when the global solvers terminate due to memory limits,
while highlighting the relevance of our approach based on global optimization methods.

%
%
%

\subsection{O--NF/PM--LMFM--III ($n \geq 4$)} In Tab.~\ref{tab:morefollowers}, we report the average computing times obtained with SCIP with employing formulations O--NF--LMFM--III and O--PM--LMFM--III.
In the time limit, we can solve NF games with up to $m=5$ for $n \leq 4$ (corresponding to $m^n\simeq 600$ different outcomes and $nm^n\simeq 2,400$ different payoffs) and up to $m=4$ for $n\leq 6$ (corresponding to about 4,000 outcomes and 24,000 payoffs). Quite interestingly, with our methods we can tackle instances of a comparable size to that of the largest instances used in~\cite{DBLP:journals/geb/PorterNS08} to evaluate a set of algorithms proposed to find a Nash equilibrium (in a single level problem), in spite of our problem being clearly harder (as it admits the former as a subproblem).
With PM games, our algorithms scale much better, allowing to find exact solutions to PM games with up to $m=10$ for $n\leq 9$ and up to $m=7$ for $n\leq 10$.

\begin{table}[h!]
\caption{Computing times (in seconds) with SCIP and O--NF/PM--LMFM--III, within a time limit of 3,600 seconds.}
\centering
\label{tab:morefollowers}
\begin{tiny}
\begin{tabular}{c|c|c|c|c}
\multicolumn{5}{c}{Normal--form games}\\
\hline
$n$ / $m$ & 2 & 3 & 4 & 5\\
\hline
3 & 0.06 & 0.20 & 0.92 & 23.79\\
 4 & 0.19 & 8.274 & 142.66 & 1304.45\\
 5 & 278.06 & 409.78 & 2016.97 & ---\\
 6 & 172.90 & 2350.95 & 2212.95 & ---
\end{tabular}
\end{tiny}
\vspace{0.1cm}
\begin{tiny}
\begin{tabular}{c|c|c|c|c|c|c}
\multicolumn{7}{c}{Polymatrix games}\\
\hline
$n$ / $m$  & 5 & 6 & 7 & 8 & 9 & 10\\
\hline
 3 & 0.24 & 2.17 & 1.87 & 7.31 & 24.45 & 194.71\\
 4 & 4.84 & 10.85 & 121.57 & 247.84 & 622.72 & 1947.54\\
 5 & 7.51 & 90.83 & 332.04 & 1982.77 & 2396.01 & 2175.29\\
 6 & 10.31 & 1169.50 & 2062.75 & --- & --- & ---
\end{tabular}
\end{tiny}
\end{table}

\subsection{O--NF/PM--LPFM and\\O--NF/PM--Implicit--Enumeration ($n = 3$)} 
\begin{figure}[!htbp]
\subfigure[O--NF--LPFM--III, SCIP]{\includegraphics[width=0.23\textwidth]{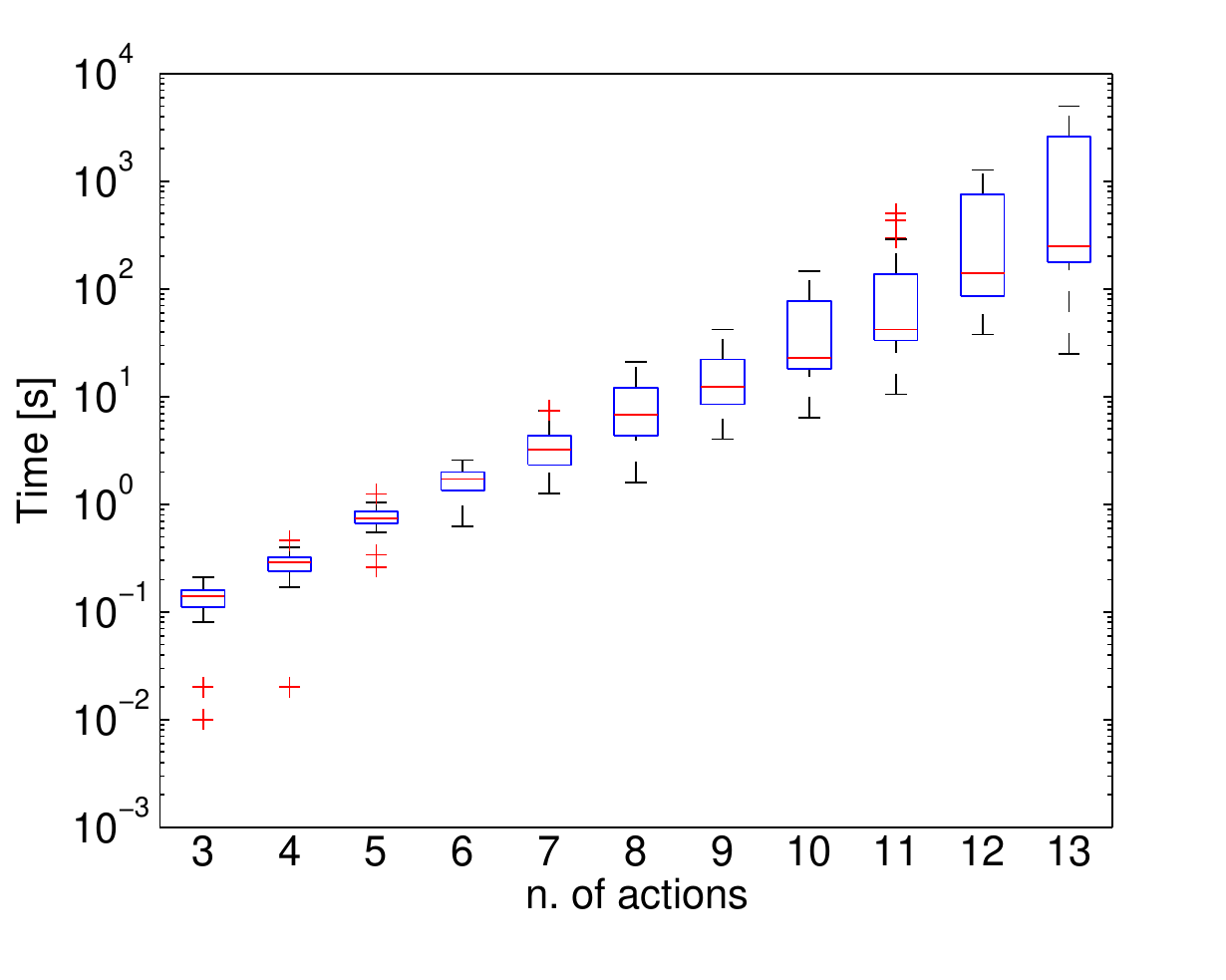}}
\subfigure[O--NF--LPFM--Impl.-Enum.]{\includegraphics[width=0.23\textwidth]{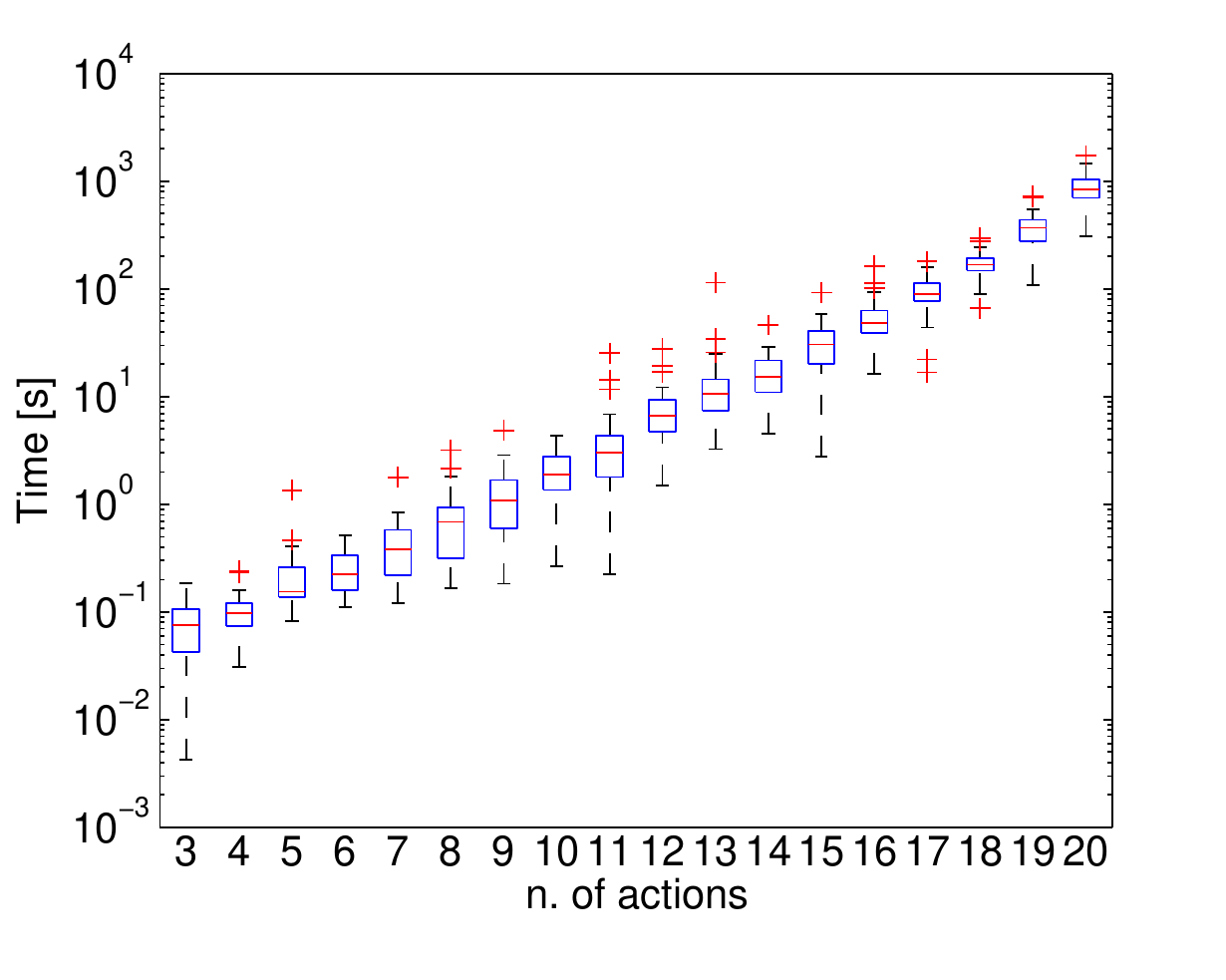}}
\subfigure[O--PM--LPFM--III, SCIP]{\includegraphics[width=0.23\textwidth]{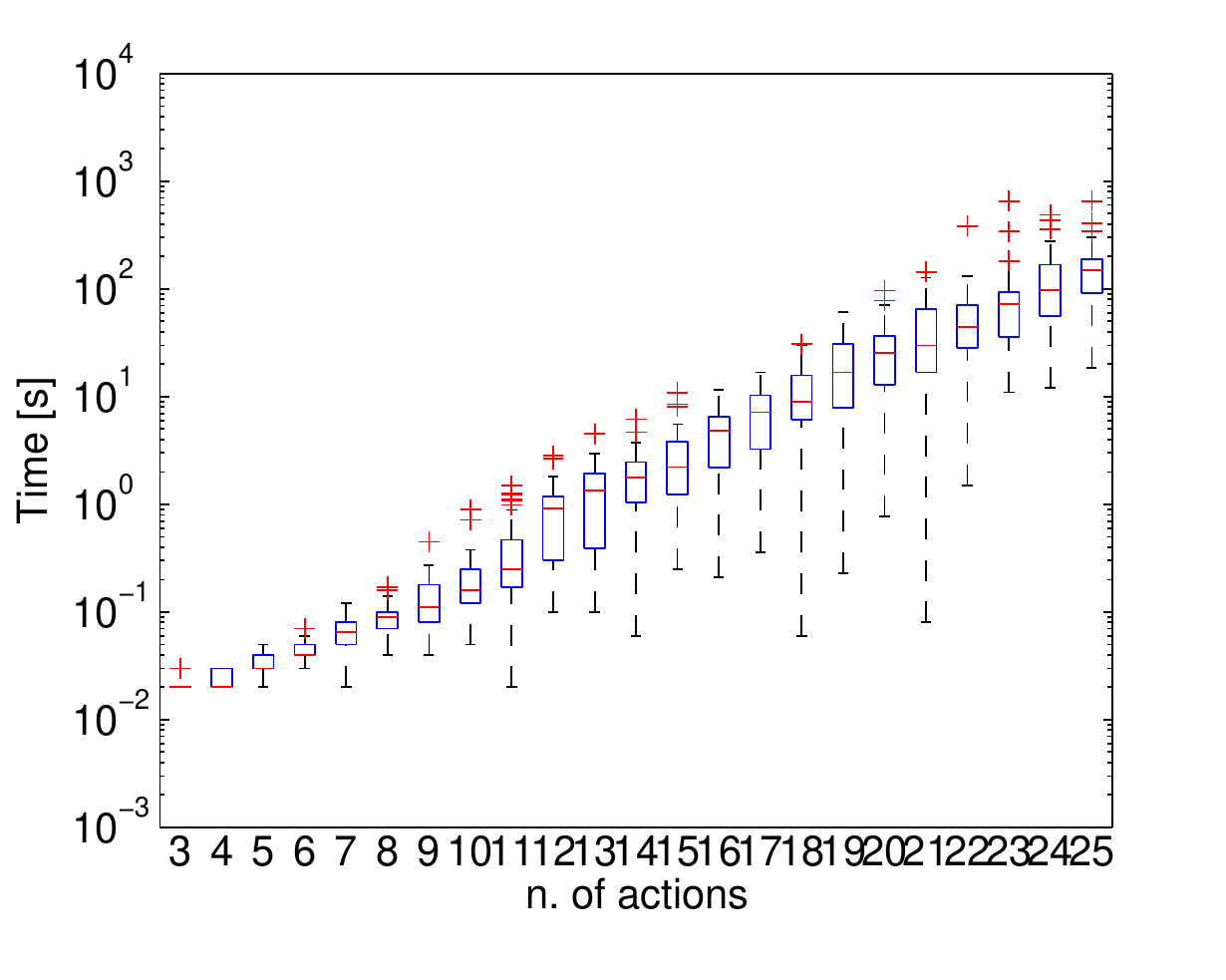}}
\subfigure[O--PM--LPFM--Impl.-Enum.]{\includegraphics[width=0.23\textwidth]{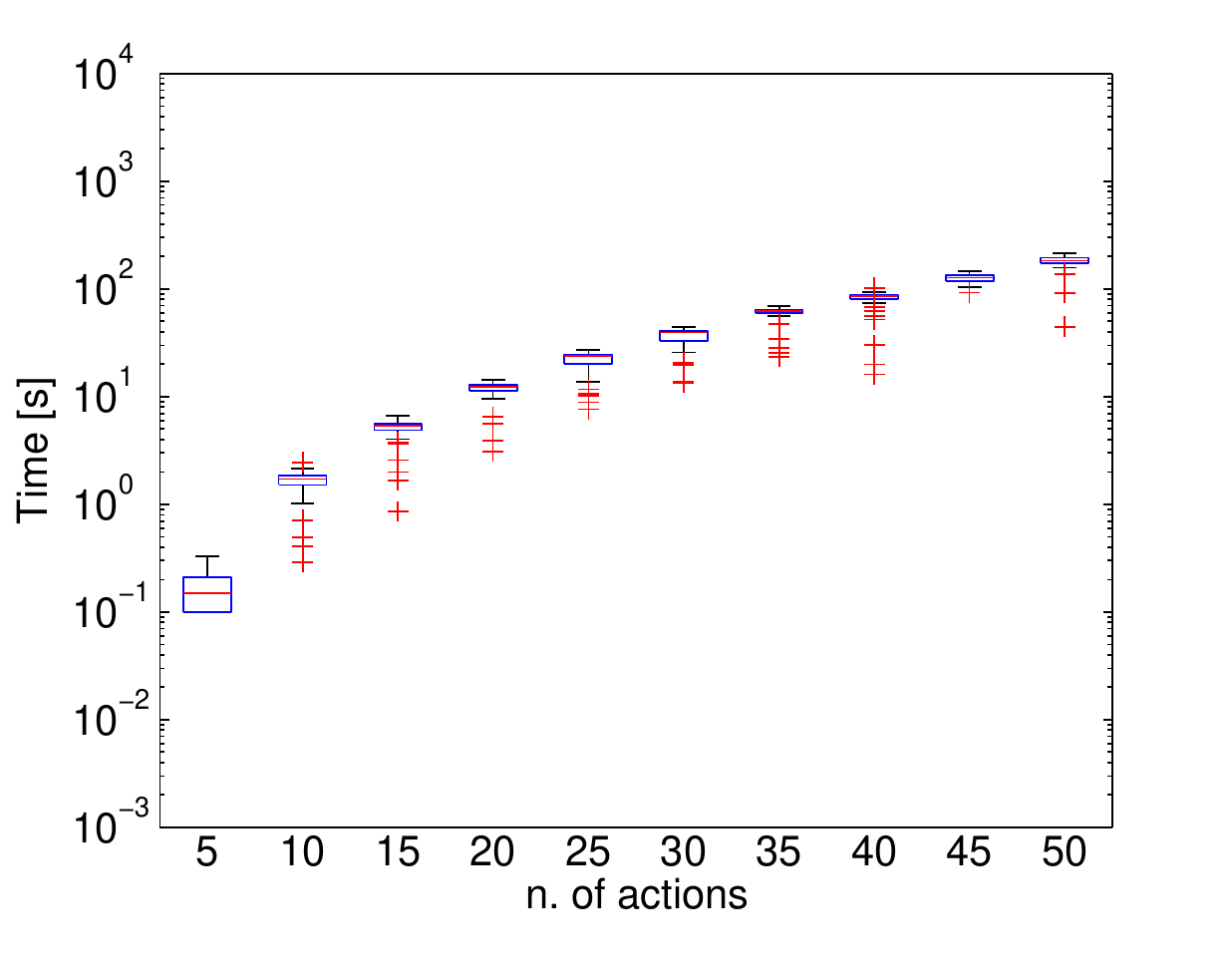}}
\caption{Computing times on NF/PM--LPFM instances with O--NF/PM--LPFM--III~(a/c) and O--NF/PM--LPFM--Implicit-Enum~(b/d), using SCIP/CPLEX.}
\label{fig:LPFM}
\end{figure}

We focus on the case where the leader is only entitled to pure strategies. We report the results only in terms of computing times obtained with SCIP for {\tt RandomGames} in Fig.~\ref{fig:LPFM}(a,b) and with CPLEX for {\tt PolymatrixGames} (for which the formulation becomes an MILP) in Fig.~\ref{fig:LPFM}(c,d). By imposing $\delta \in \{0,1\}^{m}$ in O--NF/PM--LPFM, the size of the largest instances solvable within the time limit increases from $m=9$ to $m=13$ in {\tt RandomGames} and from $m=15$ to $m=25$ for {\tt PolymatrixGames}.
For both {\tt RandomGames} and {\tt PolymatrixGames}, a dramatic performance improvement is obtained with O--NF/PM--LPFM--Implicit--Enumeration: with it, the size of the largest instance that we can solve increases from $m=13$ to $m=20$ for {\tt RandomGames} and from $m=25$ to $m=50$ for {\tt PolymatrixGames}. As expected, the computing times for {\tt PolymatrixGames} are much smaller (due to only requiring the solution of MILPs at each step), allowing us to solve to optimality much larger instances.

\subsection{O/P--NF/PM--LMFM--BlackBox} 
\begin{figure}[!htbp]
\subfigure[Average objective]{\includegraphics[width=0.23\textwidth]{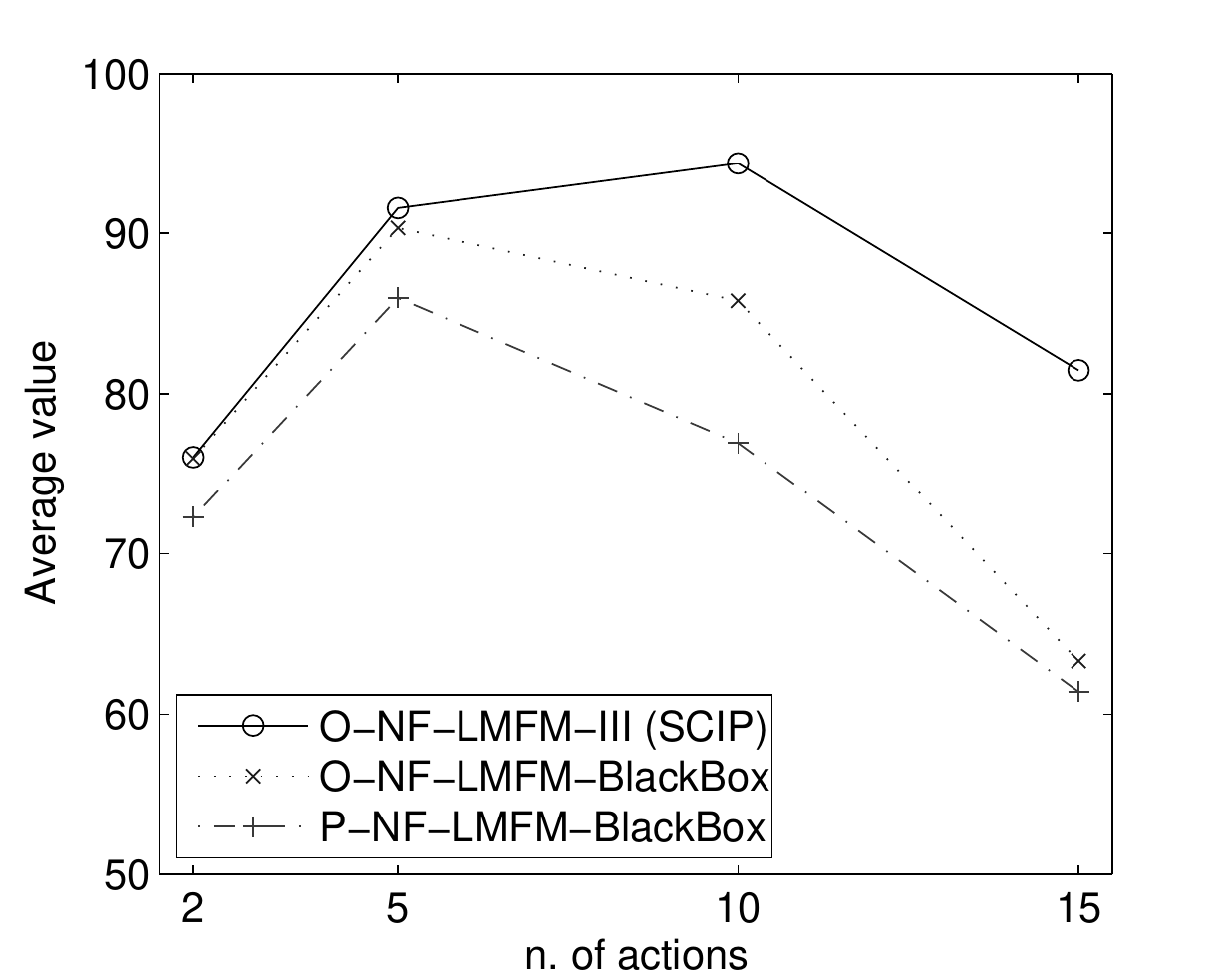}}
\subfigure[Average oracle time]{\includegraphics[width=0.23\textwidth]{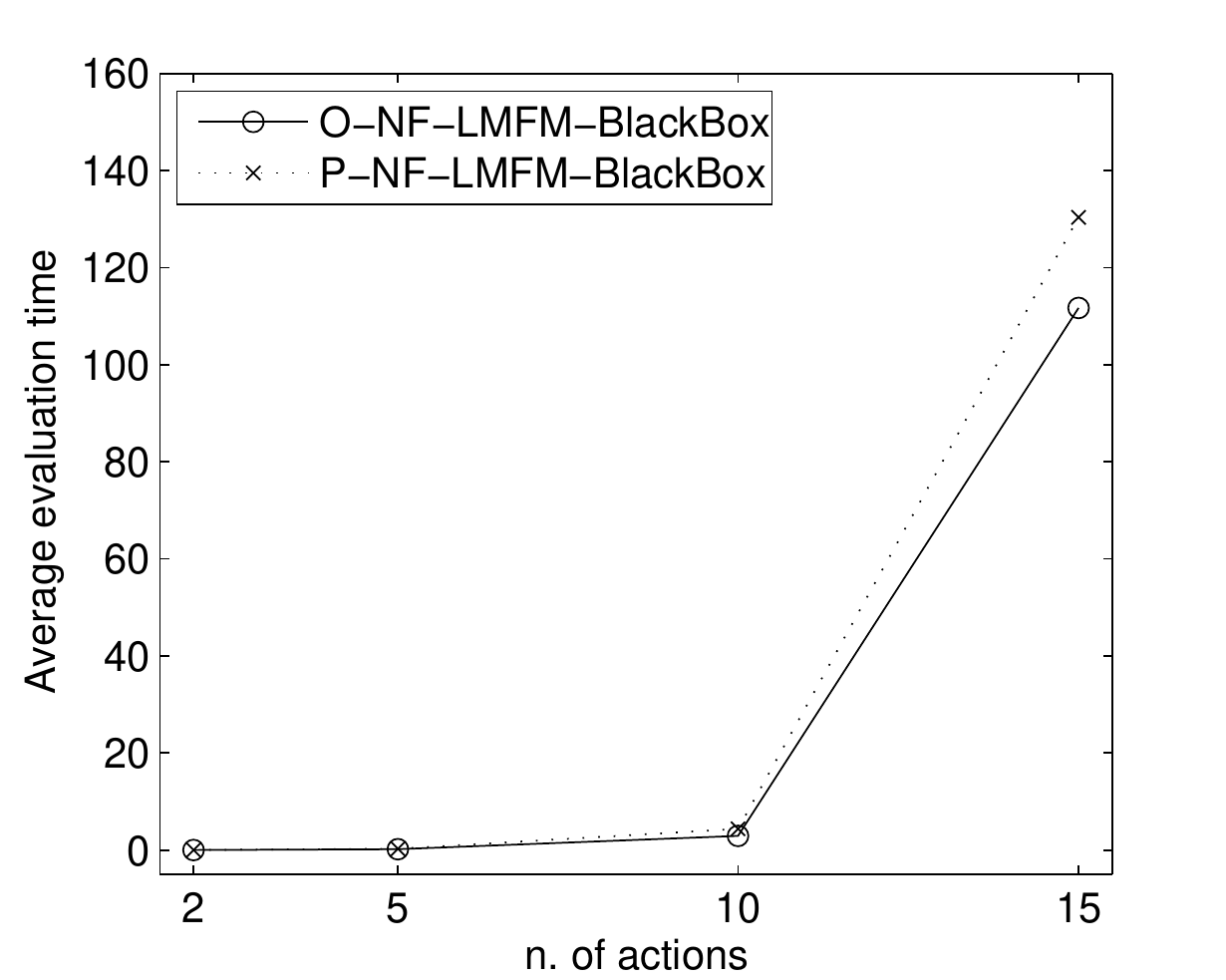}}
\caption{Performance of the Black Box approach for O/P--NF--LMFM compared to O--NF--LMFM--III.}
\label{fig:blackbox}
\end{figure}

We first consider the optimistic case for NF games, comparing, in the time limit, O--NF--LMFM--III, solved with SCIP, to O--NF--LMFM--BlackBox. For $m \leq 10$, see Fig.~\ref{fig:blackbox}(a), we observe, on average, that the black box method yields solutions within 90\% of the optimal ones found with SCIP. This suggests that the method might be sufficiently accurate. Unfortunately, for $m \geq 20$, the burden of calling SCIP to solve the oracle formulation becomes too large, see Fig.~\ref{fig:blackbox}(b), making the black box algorithm based on RBFOpt highly impractical. Notice that the method allows us to produce feasible solution also in the pessimistic case, although we cannot verify the quality of the solutions it yields. An interesting result, see Fig.~\ref{fig:blackbox}(a), concerns the gap between the utility of the leader at an optimistic LFE--N or at a pessimistic LFE--N. On the instances solved to optimality ($m \le 5$), where we can verify the quality of the black box solutions,  we see that the gap is rather small, suggesting that, in {\tt RandomGames}, the leader could be in the position to force the followers to play a strategy which provides him with a utility not dramatically smaller than that which he would obtain in an optimistic LFE--N. 

\begin{figure}[!htbp]
\subfigure[Average objective]{\includegraphics[width=0.23\textwidth]{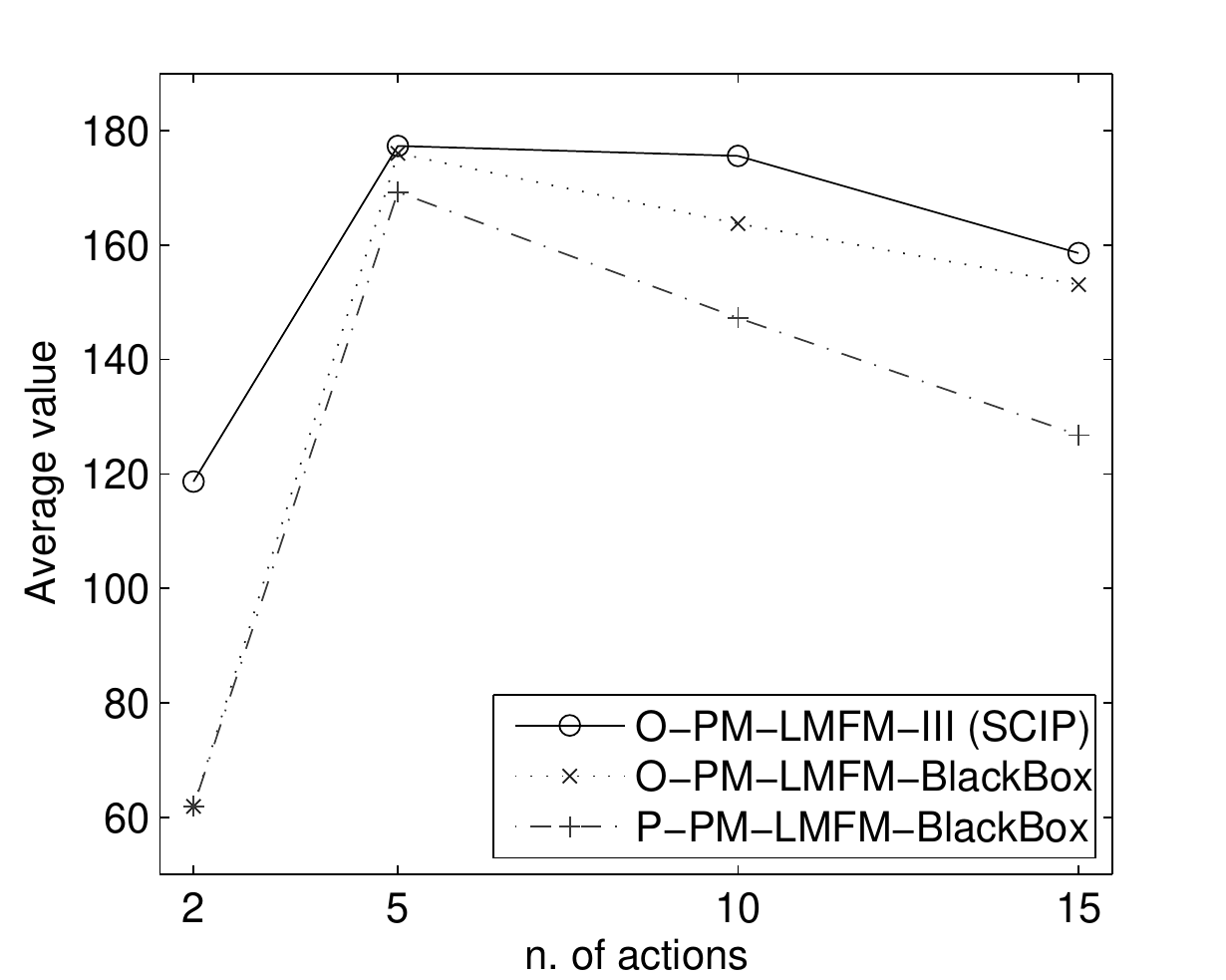}}
\subfigure[Average oracle time]{\includegraphics[width=0.23\textwidth]{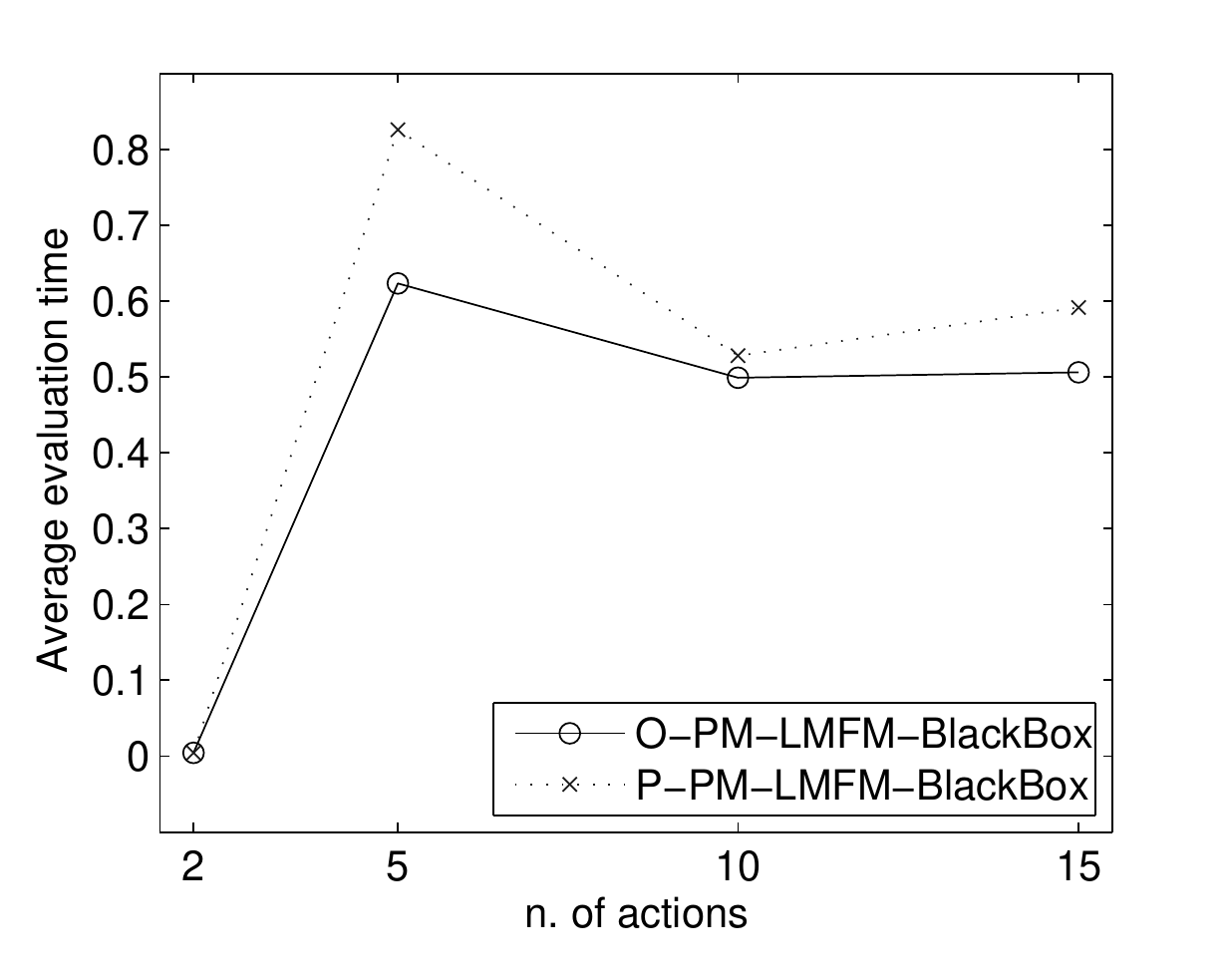}}
\caption{Performance of the Black Box approach for O/P--PM--LMFM compared to O--PM--LMFM--III.}
\label{fig:blackbox-PM}
\end{figure}

For the sake of completeness we also report, in Fig.~\ref{fig:blackbox-PM}, the analogous results obtained with polymatrix games. We compare, in the time limit, O--PM--LMFM--III, solved with SCIP, to O--PM--LMFM--BlackBox. Differently from the NF case, Figure~\ref{fig:blackbox-PM}(b) shows that, for PM games, the computing time needed to solve the oracle formulation, which is an MILP in the PM case, is much smaller and scales much better with $m$. Except for the case of $m=2$, Figure~\ref{fig:blackbox-PM}(a) allows us to draw comparable conclusions to those we drew for the NF case, with the leader achieving, in the pessimistic case, solutions not too far away, w.r.t. his utility, from the corresponding optimistic ones.

\section{Conclusions and future work}
We have provided the first computational study of game--theoretic leader--follower situations where multiple followers play a Nash equilibrium once the leader has committed to a strategy. We have provided different algorithms and mathematical programming formulations to find an equilibrium for the optimistic case where the followers maximize the leader's utility, as well as a heuristic black box method for the pessimistic case. We have conducted a thorough experimental evaluation of the different mathematical programs by means of various optimization solvers, aimed at identifying the best solver and formulation pair for both normal--form and polymatrix games. Our experiments suggest that global optimization solvers can be used as effective empirical approximation algorithms, providing a good optimality gap even for large games.

Among the challenging issues that we are interested to address in the future, we mention the design of algorithms to find an equilibrium when the followers play either a strong Nash equilibrium, a strong correlated equilibrium, or a solution concept defined in cooperative game theory.

\appendix

\section{Computational complexity}

We tackle two LFE--N problems, the optimistic version:
\begin{quote} {\bf O--LFE--N}: Given an $n$--agent game with $n\geq 3$, find a strategy vector $\delta$ for the leader such that, after committing, the NE in the followers' game, which maximizes the leader's utility given $\delta$, yields the largest value for all possible values of $\delta$.
\end{quote}
and the pessimistic version:
\begin{quote} {\bf P--LFE--N}: Given an $n$--agent game with $n\geq 3$, find a strategy vector $\delta$ for the leader such that, after committing, the NE in the followers' game, which minimizes the leader's utility given $\delta$, yields the largest value for all possible values of $\delta$.
\end{quote}

We can state the following result.
\begin{proposition}\label{apxhardness}
The problem of computing both an O--LFE--N or a P--LFE--N is $\mathcal{FNP}$--hard and it is not in Poly--$\mathcal{APX}$ unless $\mathcal{P}=\mathcal{NP}$, even when the game is polymatrix.
\end{proposition}
\emph{Proof.}
In~\cite{sandholmComplexiy2008}, the authors show that for any SAT instance it is possible to build a symmetric 2--player game $(U_1,U_2)$ such that: i) there is a pure--strategy NE in which both players play their last action providing each player with a utility of $\epsilon>0$ where $\epsilon$ is arbitrarily small, and ii) there are mixed--strategy NEs if and only the SAT instance admits a YES solution and these NEs provide each player with a utility of $\Theta(m)$, where $m$ is the number of actions. This shows that the problem of deciding whether such games admit an NE providing the players with a utility strictly larger than $\epsilon$ is $\mathcal{NP}$--hard and and finding a NE maximizing the social welfare is not in $\mathcal{APX}$. The result can be strengthen by setting $\epsilon = 2^m$ (notice that these instances can be represented with a number of bits linear in $m$), showing that no better approximation ratio than $1/2^m$ can be found in polynomial time and, therefore, that the problem is not in Poly--$\mathcal{APX}$.

We now extend this result to O/P--LFE--N.
Given $(U_1,U_2)$ with $m$ actions per player as defined in~\cite{sandholmComplexiy2008}, construct a 3--player leader--follower polymatrix game where:
\begin{itemize}
\item the leader $\ell$ only has one action and his utility matrices are $U_{\ell f_1} = U_{\ell f_2} = [1,1,\ldots,1,1/2^m]$;
\item player $f_1$'s utility matrices are $U_{f_1\ell}=\mathbf{0}$ and $U_{f_1f_2}=U_{1}$;
\item player $f_2$'s utility matrices are $U_{f_2\ell}=\mathbf{0}$ and $U_{f_2f_1}=U_{2}$.
\end{itemize}
It can be easily seen that approximating in polynomial time the $\ell$'s expected utility with an approximation ratio better than $1/2^m$ provides us with an algorithm to decide in polynomial time whether there is an NE in $(U_1,U_2)$ providing each player with a utility strictly larger than $\epsilon$. This shows that O--LFE--N is not in Poly--$\mathcal{APX}$ unless $\mathcal{P}=\mathcal{NP}$ even in polymatrix games (which are a special case of normal form games).

For P--LFE--N, where the followers play an NE which minimizes the leader's utility, the reduction is the same, except for letting $U_{\ell f_1} = U_{\ell f_2} = [1/2^m,1/2^m,\ldots,1/2^m,1]$.\QEDB
\medskip

Furthermore, we show that
deciding whether one of the leader's actions can be safely discarded is a hard problem, thus showing that efficient dominance--like techniques are inapplicable.
\begin{proposition}
In a leader--follower game in which the followers play the best (for the leader) Nash equilibrium, deciding whether or not an action of the leader is played with strictly positive probability at the optimistic LFE--N is $\mathcal{NP}$--hard.
\end{proposition}
\emph{Proof}.
Given a symmetric 2--player $(U_1,U_2)$ with $m$ actions of the form used in the reduction in~\cite{sandholmComplexiy2008}, we build a 3--player game $(U_{\ell},U_{f_1},U_{f_2})$ in which:
\begin{itemize}
\item $\ell$ has two actions, while $f_1$ and $f_2$ have $m$ actions;
\item for the first action of $\ell$, the payoffs of all the players are $1/4$;
\item for the second action of $\ell$, the payoffs of $f_1$ and $f_2$ are those in $(U_1,U_2)$, while the payoffs of $\ell$ are 1 for all the actions of $f_1$ and $f_2$ except for the combination composed of the last action of $f_1$ and the last action of $f_2$, in which the payoff of $\ell$ is 0.
\end{itemize}
For the properties of such games, see the proof of Proposition 1.

We show that the first action of $\ell$ can be safely discarded from the NF game $(U_{\ell},U_{f_1},U_{f_2})$ if and only if the game $(U_1,U_2)$ admits a mixed NE. Therefore, the problem of deciding whether the first action of $\ell$ can be discarded is $\mathcal{NP}$--hard.
If $\ell$ plays his first action, he receives a utility of $1/4$. If $\ell$ plays his second action, the followers play the best NE for the leader. It can  be: either i) the pure--strategy NE in which both play their last action providing $\ell$ with a utility of 0 or, ii) it if exists, the mixed--strategy NE providing $\ell$ with a utility of 1. For any mixed strategy of $\ell$, the behavior of the followers does not change w.r.t. the case in which $\ell$ plays purely his second action. This is because, when $\ell$ randomizes between his two actions, the utility of the followers $f_1$ and $f_2$ is an affine transformation (with positive coefficient) of $U_1$ and $U_2$
and, therefore, they play exactly as in the case $\ell$ plays purely his second action. Thus, it can be easily observed that, at an optimistic LFE--N, $\ell$ plays a pure strategy, playing his first action when $(U_1,U_2)$ does not admit the mixed--strategy NE and his second action otherwise. Thus, if there is the mixed--strategy  NE in $(U_1,U_2)$, then the first action of $\ell$ can be safely discarded, while it cannot be otherwise.  The claim follows.
\hfill$\Box$

\bibliographystyle{abbrv}

\bibliography{citations}

\begin{thebibliography}{10}

\bibitem{an2011guards}
B.~An, J.~Pita, E.~Shieh, M.~Tambe, C.~Kiekintveld, and J.~Marecki.
\newblock Guards and {Protect}: Next generation applications of security games.
\newblock {\em ACM SIGecom Exchanges}, 10(1):31--34, 2011.

\bibitem{DBLP:conf/aaai/ConitzerK11}
V.~Conitzer and D.~Korzhyk.
\newblock Commitment to correlated strategies.
\newblock In {\em Proceedings of the Twenty--Fifth {AAAI} Conference on
  Artificial Intelligence {AAAI}}, 2011.

\bibitem{Conitzer:2006:COS:1134707.1134717}
V.~Conitzer and T.~Sandholm.
\newblock Computing the optimal strategy to commit to.
\newblock In {\em ACM Conference on Electronic Commerce}, pages 82--90, 2006.

\bibitem{sandholmComplexiy2008}
V.~Conitzer and T.~Sandholm.
\newblock New complexity results about {Nash} equilibria.
\newblock {\em GAME ECON BEHAV}, 63(2):621--641, 2008.

\bibitem{blackboxoptimization}
A.~Costa, G.~Nannicini, T.~Schroepfer, and T.~Wortmann.
\newblock Black--box optimization of lighting simulation in architectural
  design.
\newblock In {\em Complex Systems Design \& Management Asia}, pages 27--39.
  2015.

\bibitem{CPLEX}
CPLEX.
\newblock http://www-03.ibm.com/software/products/en/ibmilogcpleoptistud/,
  2014.

\bibitem{dempe2003bilevel}
S.~Dempe.
\newblock {\em Bilevel programming: A survey}.
\newblock Dekan der Fak. f{\"u}r Mathematik und Informatik, 2003.

\bibitem{Gill97snopt:an}
P.~E. Gill, W.~Murray, and M.~A. Saunders.
\newblock Snopt: An sqp algorithm for large-scale constrained optimization.
\newblock {\em SIAM Journal on Optimization}, 12:979--1006, 1997.

\bibitem{DBLP:conf/atal/KiekintveldJTPOT09}
C.~Kiekintveld, M.~Jain, J.~Tsai, J.~Pita, F.~Ord{\'{o}}{\~{n}}ez, and
  M.~Tambe.
\newblock Computing optimal randomized resource allocations for massive
  security games.
\newblock In {\em International Joint Conference on Autonomous Agents and
  Multiagent Systems {(AAMAS)}}, pages 689--696, 2009.

\bibitem{mccormick1976computability}
G.~McCormick.
\newblock {Computability of global solutions to factorable nonconvex programs:
  Part I -- Convex underestimating problems}.
\newblock {\em Math. Program.}, 10(1):147--175, 1976.

\bibitem{gamut}
E.~Nudelman, J.~Wortman, K.~Leyton-Brown, and Y.~Shoham.
\newblock Run the {GAMUT}: A comprehensive approach to evaluating
  game--theoretic algorithms.
\newblock In {\em AAMAS}, pages 880--887, 2004.

\bibitem{DBLP:journals/geb/PorterNS08}
R.~Porter, E.~Nudelman, and Y.~Shoham.
\newblock Simple search methods for finding a nash equilibrium.
\newblock {\em Games and Economic Behavior}, 63(2):642--662, 2008.

\bibitem{BARON}
N.~V. Sahinidis.
\newblock {\em {BARON 14.3.1: Global Optimization of Mixed--Integer Nonlinear
  Programs, {\em User's Manual}}}, 2014.

\bibitem{sandholmgilpinconitzer2005}
T.~Sandholm, A.~Gilpin, and V.~Conitzer.
\newblock Mixed--integer programming methods for finding {Nash} equilibria.
\newblock In {\em AAAI}, pages 495--501, 2005.

\bibitem{SCIP}
SCIP.
\newblock Scip (solving constraint integer programs) optimization suite, 2014.

\bibitem{shoham-book}
Y.~Shoham and K.~Leyton-Brown.
\newblock {\em Multiagent Systems: Algorithmic, Game Theoretic and Logical
  Foundations}.
\newblock Cambridge University Press, 2008.

\bibitem{leaderfollower}
B.~von Stengel and S.~Zamir.
\newblock Leadership games with convex strategy sets.
\newblock {\em Game and Economic Behavior}, 69:446--457, 2010.

\bibitem{polymatrixref}
E.~B. Yanovskaya.
\newblock Equilibrium points in polymatrix games.
\newblock {\em Lithuanian Mathematical Journal}, 8:381--384, 1968.

\end{thebibliography}

\end{document}